\documentclass[11pt]{article}

\usepackage{epsfig}
\usepackage{latexsym}
\usepackage{graphicx}
\usepackage{color}
\usepackage{amsmath,amssymb}
\usepackage{cite}

\makeatletter
\def\section{\@startsection {section}{1}{\z@}{-3.5ex plus -1ex minus -.2ex}{2.3ex plus .2ex}{\large\bf}}
\def\subsection{\@startsection{subsection}{2}{\z@}{-3.25ex plus -1ex
minus -.2ex}{1.5ex plus .2ex}{\normalsize\bf}}
\makeatother
\newcommand{\captionfonts}{\small}
\makeatletter  
\long\def\@makecaption#1#2{%
  \vskip\abovecaptionskip
  \sbox\@tempboxa{{\captionfonts #1: #2}}%
  \ifdim \wd\@tempboxa >\hsize
    {\captionfonts #1: #2\par}
  \else
    \hbox to\hsize{\hfil\box\@tempboxa\hfil}%
  \fi
  \vskip\belowcaptionskip}
\makeatother   

\catcode`@=11
\def\marginnote#1{}
\newcount\hour
\newcount\minute
\newtoks\amorpm
\hour=\time\divide\hour
by60
\minute=\time{\multiply\hour by60 \global\advance\minute
by-\hour}
\edef\standardtime{{\ifnum\hour<12 \global\amorpm={am}
\else\global\amorpm={pm}\advance\hour by-12 \fi
 \ifnum\hour=0
\hour=12 \fi
 \number\hour:\ifnum\minute<10
0\fi\number\minute\the\amorpm}}
\edef\militarytime{\number\hour:\ifnum\minute<10
0\fi\number\minute}
\def\draftlabel#1{{\@bsphack\if@filesw
{\let\thepage\relax
 \xdef\@gtempa{\write\@auxout{\string
\newlabel{#1}{{\@currentlabel}{\thepage}}}}}\@gtempa
 \if@nobreak
\ifvmode\nobreak\fi\fi\fi\@esphack}
\gdef\@eqnlabel{#1}}
\def\@eqnlabel{}
\def\@vacuum{}
\def\draftmarginnote#1{\marginpar{\raggedright\scriptsize\tt#1}}
\def\draft{\oddsidemargin
0.0truein
 \def\@oddfoot{\sl preliminary draft \hfil
\rm\thepage\hfil\sl\today\quad\militarytime}
 \let\@evenfoot\@oddfoot
\overfullrule 3pt
 \let\label=\draftlabel
\let\marginnote=\draftmarginnote
\def\@eqnnum{(\theequation)\rlap{\kern\marginparsep\tt\@eqnlabel}
\global\let\@eqnlabel\@vacuum}
}
\catcode`@=12

\newcommand{\beq}{\begin{eqnarray}}
\newcommand{\eeq}{\end{eqnarray}}
\def\mchij{m_{\tilde{\chi}^+_j}}
\def\mchi01{m_{\tilde{\chi}^0_1}}
\def\mst1{m_{\tilde{t}_1}}
\def\msc1{m_{\tilde{c}_L}}
\def\mn1{m_{\tilde{\chi}_1^0}}
\def\mdk{m_{d_k}}

\def\mdki{m_{\tilde{d}_{ki}}}
\def\acdj1{a_{j1}^{\tilde{c} d_k}}
\def\atdj1{a_{j1}^{\tilde{t} d_k}}
\def\bcdj1{b_{j1}^{\tilde{c} d_k}}
\def\btdj1{b_{j1}^{\tilde{t} d_k}}

\newcommand{\s}{\newline \vspace*{-3.5mm}}

\begin{document}

\thispagestyle{empty}

\begin{center}
\hfill KA-TP-06-2011

\begin{center}

\vspace{1.7cm}

{\LARGE\bf Light Stop Decay in the MSSM with Minimal Flavour Violation}
\end{center}

\vspace{1.4cm}

{\bf M. M\"uhlleitner$^{\,a}$} and {\bf E. Popenda$^{\,a}$} \\

\vspace{1.2cm}

${}^a\!\!$
{\em {Institut f\"ur Theoretische Physik, Karlsruhe Institute of Technology, 76128 Karlsruhe, Germany}
}\\

\end{center}

\vspace{0.8cm}

\centerline{\bf Abstract}
\vspace{2 mm}
\begin{quote}
\small
In supersymmetric scenarios with a light stop particle $\tilde{t}_1$
and a small mass difference to the lightest supersymmetric particle
(LSP) assumed to be the lightest neutralino, the flavour changing
neutral current decay $\tilde{t}_1 \to c \tilde{\chi}_1^0$ can be the
dominant decay channel and can exceed the four-body   stop decay for
certain parameter values. In the framework of Minimal Flavour
Violation (MFV) this decay is CKM-suppressed, thus inducing long stop
lifetimes. Stop decay length measurements at the LHC can then be exploited
to test models with minimal flavour breaking through Standard Model
Yukawa couplings. The decay width has been given some time ago by an
approximate formula, which takes into account the leading logarithms
of the MFV scale. In this paper we calculate the exact one-loop decay
width in the framework of MFV. The comparison with the approximate
result exhibits deviations of the order of 10\% for large MFV scales
due to the neglected non-logarithmic terms in the approximate decay
formula.  The difference in the branching ratios is negligible. 
The large logarithms have to be resummed. The
resummation is performed by the solution of the renormalization group
equations. The comparison of the exact one-loop result and the tree
level flavour changing neutral current decay, which incorporates the
resummed logarithms, demonstrates that the resummation effects are
important and should be taken into account.  
\end{quote}

\newpage

\section{\label{sec:Intro} Introduction}
The Standard Model (SM) provides a very successful effective theory of
particle interactions which is in excellent agreement with electroweak
precision data. Furthermore, remarkable consistency and precision
tests have been made in the sector of quark flavour violation.
These tests and limits on flavour changing neutral currents (FCNC) from
$K,D$ and $B$ studies put strong constraints on possible New
Physics beyond the SM \cite{Grossman:2009dw}. They forbid a generic
flavour structure of New Physics at the TeV scale and raise the
question why contributions from New Physics at $\sim 1$ TeV are
strongly suppressed. A solution to this
New Physics Flavour Puzzle is provided by the framework of Minimal
Flavour Violation (MFV)
\cite{Chivukula:1987fw,Buras:2000dm,D'Ambrosio:2002ex,Bobeth:2005ck}.  
It requires all sources of flavour and CP-violation to be given by the
SM structure of the Yukawa couplings. Flavour mixing in models of New
Physics is then always proportional to the off-diagonal elements of the
Cabibbo-Kobayashi-Maskawa (CKM) matrix \cite{Cabibbo:1963yz}.
In particular, the mixing of the third generation squarks with the first and
second generation squarks is highly suppressed by small CKM quark
mixing angles. \s

Since the flavour structure of New Physics at the TeV scale must be
non-generic, flavour measurements provide a good probe of New
Physics. One of the best studied examples is given by
supersymmetry (SUSY). In the context of flavour violation the
phenomenology of the light stop partner $\tilde{t}_1$ is especially
interesting. In most SUSY models a light stop quark mass 
arises naturally. Due to the large top Yukawa coupling the mixing
between the weak eigenstates $\tilde{t}_L$ and $\tilde{t}_R$ leads to
a large mass splitting between the stop mass eigenstates. Furthermore,
the large top Yukawa coupling in general drives the stop mass via the
renormalization group equation (RGE) running to smaller masses, even
if all squarks have a common mass at the SUSY breaking scale. In
scenarios with a light stop which predominantly decays into a
charm quark and the lightest neutralino, assumed to be the
lightest supersymmetric particle (LSP), $\tilde{t}_1 \to c
\tilde{\chi}^0_1$, squark flavour violation can be tested by
exploiting stop decay length measurements \cite{Hiller:2008wp}. It has been
shown, that in these scenarios light stops can be discovered at the
LHC \cite{Han:2003qe}.  Finally, a light stop is also  favoured by baryogenesis
which requires a stop mass with about the top mass value or less for
successful electroweak baryogenesis within the context of the Minimal
Supersymmetric Extension of the Standard Model (MSSM) \cite{EWBG}. \s 

Minimal Flavour Violation naturally arises in supergravity models
which provide flavour-in\-de\-pend\-ent scalar mass terms at a high
scale like the Planck scale $M_{P}$.  Some time ago the authors of
Ref. \cite{Hikasa:1987db}  have provided an approximate formula for
the flavour changing neutral current (FCNC) 
decay $\tilde{t}_1 \to c \tilde{\chi}_1^0$ by starting from a
vanishing tree level $\tilde{t}_1-c-\tilde{\chi}_1^0$ coupling at the
Planck scale.  The decay then proceeds via one-loop diagrams. The
non-vanishing divergences, which are due to scalar self-mass diagrams,
have been subtracted by a soft counterterm at the Planck scale so that
a large logarithm $\ln(M_P^2/M_W^2)$ remains at the weak scale chosen
to be $M_W$. The authors argue that in view of this large logarithm,
the  remaining non-logarithmic part of the one-loop diagrams can hence
safely be neglected so that their result for the decay width takes a
rather simple form. \s  

In this work, we perform the complete one-loop calculation of the
$\tilde{t}_1 \to c\tilde{\chi}_1^0$ decay in the framework of MFV. We
perform the full renormalization program and keep the finite
non-logarithmic terms arising from the loop integrals.
This allows us to study the importance of the neglected
non-logarithmic pieces in the formula given by Ref.\cite{Hikasa:1987db}. \s

To get a reliable result, the appearing large logarithms should be
resummed. In the renormalization group approach this corresponds to
solving the renormalization group equations for the scalar soft SUSY
breaking squark masses. As has been pointed out in
\cite{D'Ambrosio:2002ex}, the hypothesis of MFV is not renormalization
group invariant. Flavour off-diagonal squark mass terms are hence
induced by the Yukawa couplings, so that the squark and quark mass
matrices cannot be diagonalized simultaneously any more and the stop
state receives some admixture from the charm squark, inducing a FCNC
between stop, charm and LSP neutralino,
$\tilde{t}_1-c-\tilde{\chi}_1^0$. From this point of view, the
logarithmic piece of our one-loop result is equivalent to the first order in the
expansion of the RGE solution for the squark-quark-neutralino coupling
in powers of $\alpha$, whereas the tree level decay calculated
with the FCNC coupling includes the resummation of the large
logarithms. The comparison of the two decay widths provides an
estimate of the importance of the resummation of the large
logarithms. \s 

The outline of our paper is as follows: In section \ref{sec:oneloop}
we present the diagrams contributing to the one-loop decay. We set up
our notation for the squark and quark sector in section
\ref{sec:notation}. The counterterms and the renormalization are
discussed in section \ref{sec:renorm}. Section \ref{sec:numerical}
contains the numerical analysis. We conclude in section
\ref{sec:summary}. In the Appendix we list our Feynman rules, the
various amplitudes contributing to the decay and we derive the FCNC
counterterm. 

\section{\label{sec:oneloop} One-loop decay} 
\begin{figure}[b]
\begin{center}
\epsfig{figure=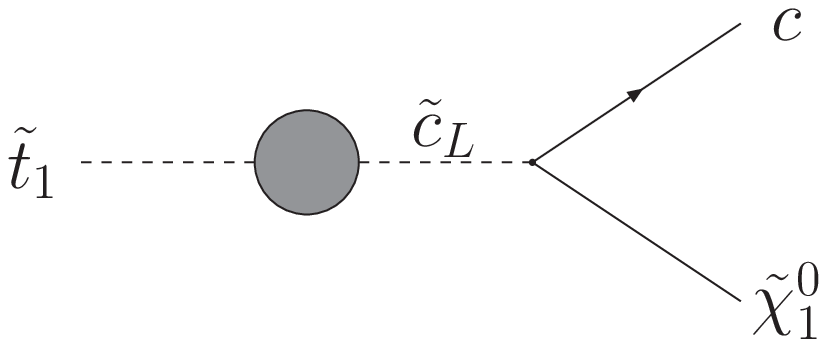,width=4.3cm,clip=}
\quad
\epsfig{figure=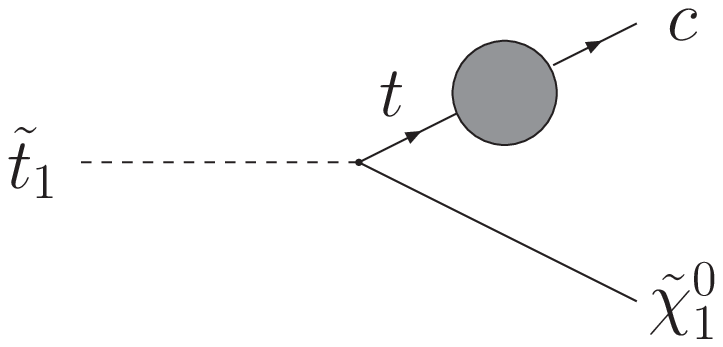,width=3.8cm,clip=}
\quad
\epsfig{figure=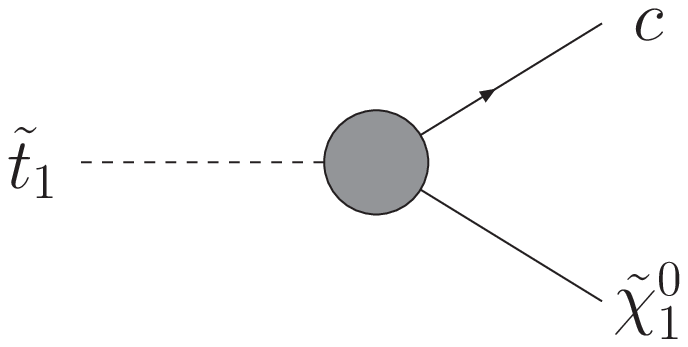,width=3.8cm,clip=}
\caption{\label{fig:gendiag} Generic diagrams contributing to the
  loop-decay $\tilde{t}_1 \to c \tilde{\chi}_1^0$.}
\end{center}
\end{figure}

\begin{figure}[t!]
\begin{tabular}{c}
\epsfig{figure=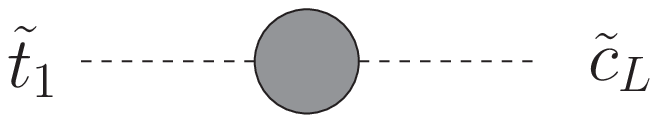,width=4.2cm}     
\end{tabular} $=$
\begin{tabular}{c}
\vspace{0.8cm}
\hspace{-0.3cm}
\epsfig{figure=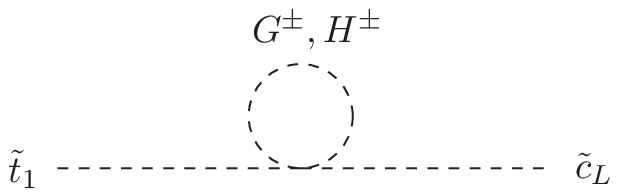,width=4.3cm}
\end{tabular} $+$
\begin{tabular}{c}
\vspace{0.8cm}
\hspace{-0.3cm}
\epsfig{figure=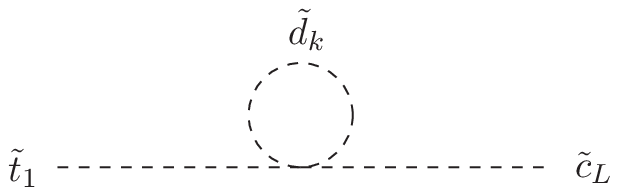,width=4.3cm}
\end{tabular} $+$\\[-0.5cm]
\begin{tabular}{c}
\epsfig{figure=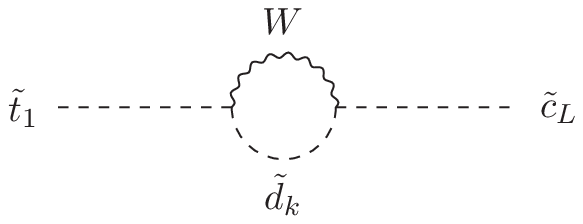,width=4.2cm}
\end{tabular} $+$
\begin{tabular}{c}
\epsfig{figure=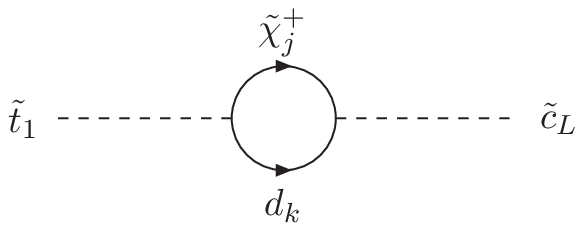,width=4.2cm}
\end{tabular} $+$
\begin{tabular}{c}
\epsfig{figure=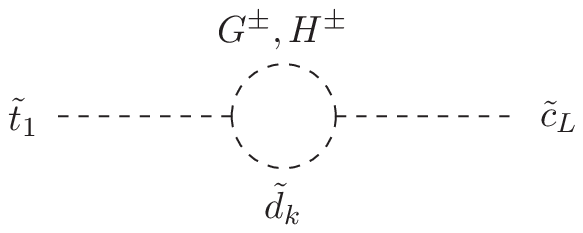,width=4.2cm}
\end{tabular}\\[0.7cm]
\begin{tabular}{c}
\epsfig{figure=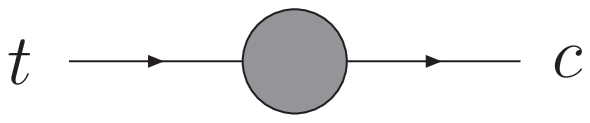,width=4cm}
\end{tabular}  $\ =\ $
\begin{tabular}{c}
\epsfig{figure=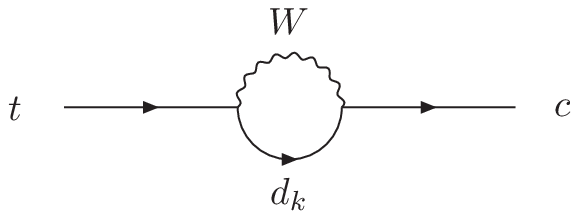,width=4cm}
\end{tabular}  $\ +\ $
\begin{tabular}{c}
\epsfig{figure=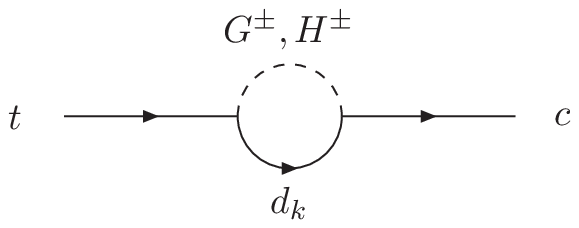,width=4cm}
\end{tabular}\\
\hspace*{4.5cm}
$\ +\ $
\begin{tabular}{c}
\hspace{-0.3cm}
\epsfig{figure=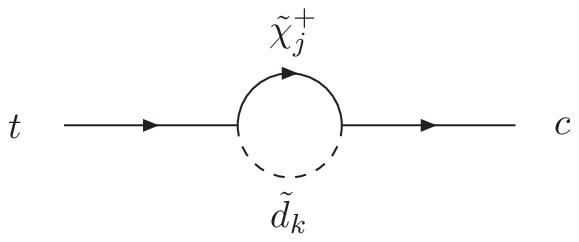,width=4.2cm}
\end{tabular}\\[0.7cm]
\begin{tabular}{c}
\epsfig{figure=vertex-proper.eps,width=4.1cm}
\end{tabular}  $\ =$
\begin{tabular}{c}
\epsfig{figure=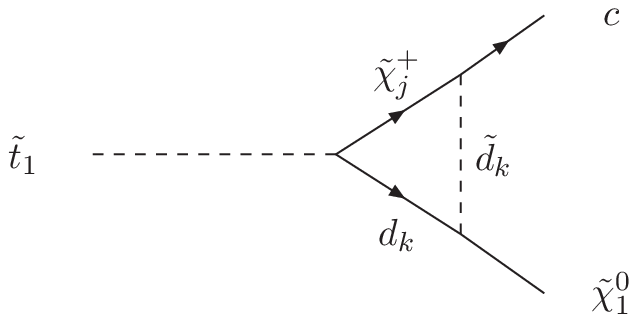,width=4.1cm}
\end{tabular}  $+$
\begin{tabular}{c}
\epsfig{figure=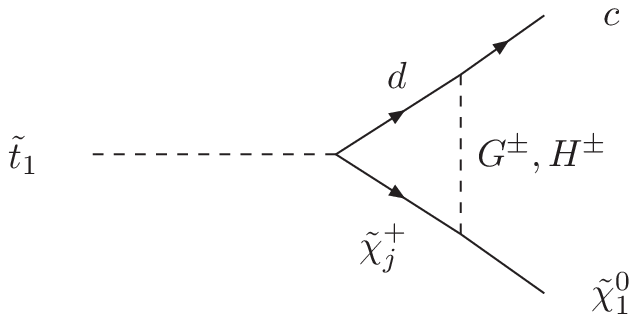,width=4.1cm}
\end{tabular}  $+$\\[0.3cm]
\begin{tabular}{c}
\epsfig{figure=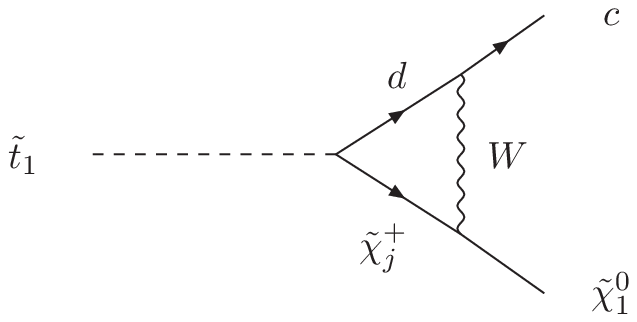,width=4.1cm}
\end{tabular}  $\ +$
\begin{tabular}{c}
\epsfig{figure=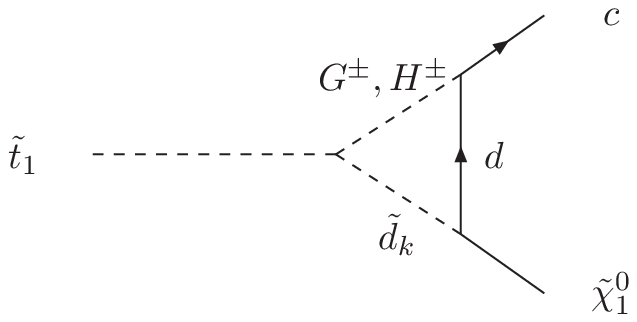,width=4.1cm}
\end{tabular}  $+$
\begin{tabular}{c}
\epsfig{figure=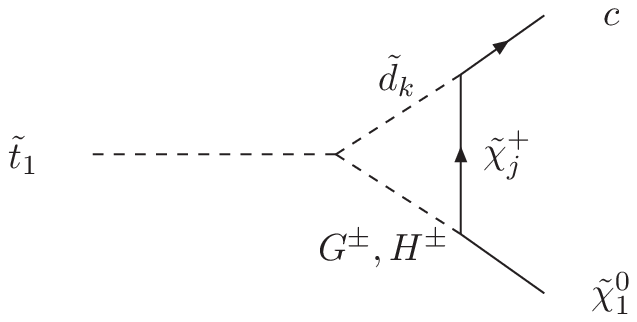,width=4.1cm}
\end{tabular}  $+$\\[0.3cm]
\begin{tabular}{c}
\epsfig{figure=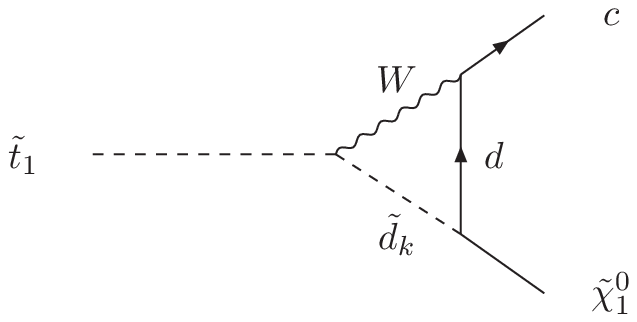,width=4.1cm}
\end{tabular}  $\ +$
\begin{tabular}{c}
\epsfig{figure=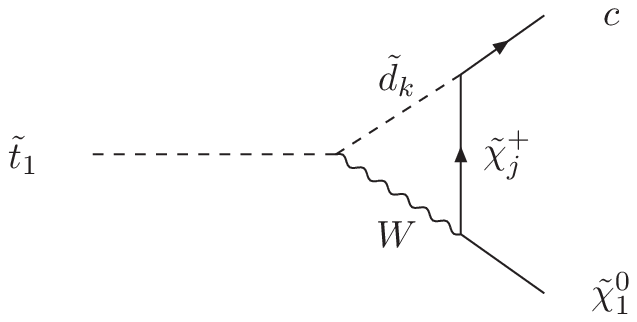,width=4.1cm}
\end{tabular}
\begin{center}
\caption{\label{fig:alldiags} Generic diagrams contributing to the squark
 and quark self-energy and the proper vertex correction.}
\end{center}
\vspace{0.4cm}
\begin{center}
\epsfig{figure=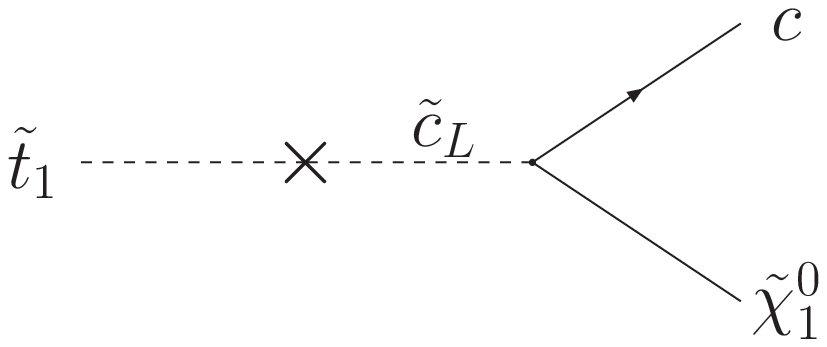,width=4.2cm,clip=}
\quad
\epsfig{figure=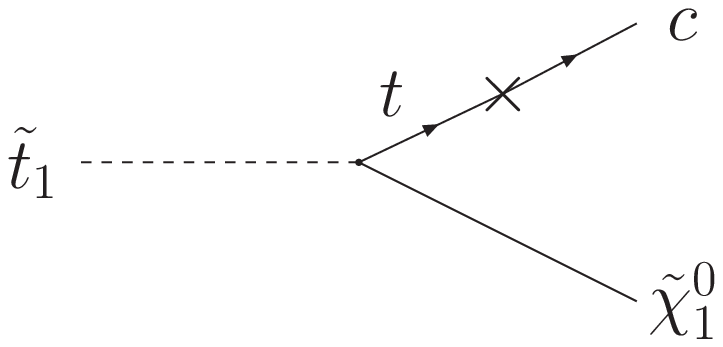,width=3.7cm,clip=}
\quad
\epsfig{figure=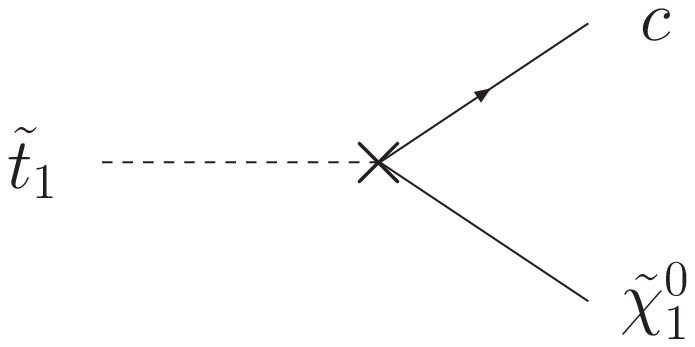,width=3.8cm,clip=}
\caption{\label{fig:counter} Counterterm diagrams.}
\end{center}
\end{figure}

We work in the framework of the MSSM with MFV so that all flavour
changing effects with quarks and squarks are controlled by the quark
Yukawa couplings and CKM mixing angles \cite{Buras:2000dm}. The decay
of the lightest stop $\tilde{t}_1$ into the lightest neutralino
$\tilde{\chi}^0_1$ and a charm quark $c$, 
\beq
\tilde{t}_1 \to c \tilde{\chi}^0_1 \, ,
\label{eq:process}
\eeq
is then mediated at the one-loop level. We consider scenarios where
the light stop $\tilde{t}_1$ is the next-to-lightest supersymmetric
particle (NLSP) and the lightest neutralino is the LSP. 
The process is built up by the stop and charm self-energies and the
vertex diagrams, {\it cf.} Fig.~\ref{fig:gendiag}. Note, that in our
calculation we set  
\beq
m_c = 0 \;.
\eeq
Therefore in the $\tilde{t}_1$ self-energies we have only
non-vanishing contributions for transitions into the left-handed charm
squark $\tilde{c}_L$. Those into right-handed scharm, $\tilde{c}_R$,
are zero for $m_c=0$. All diagrams are 
mediated by charged current loops. The various diagrams which
contribute are depicted in Fig.~\ref{fig:alldiags}. The self-energies and vertex 
corrections are divergent and have to be renormalized. The
counterterms for the squark and quark self-energies and for the vertex
renormalization are shown in Fig.~\ref{fig:counter}. 
The FCNC
vertex does not arise at tree level. Its occurrence as counterterm at
one-loop level is due to the fact that MFV is not  RGE-invariant,
since the weak interactions affect the squark and quark mass matrices
differently \cite{Donoghue:1983mx}. Their simultaneous diagonalization
cannot be maintained at higher orders so that it can consistently be
imposed at a single scale only, called $\mu_{\mbox{\scriptsize MFV}}$
in the following. \s

For the calculation of the stop decay process we define an effective
interaction vertex
\beq
T \equiv g \, \bar{u}_c (k_2) \, (F_L {\cal P}_L + F_R {\cal P}_R)
\, v_{\tilde{\chi}_1^0} (k_1) \;,
\label{eq:effvertex}
\eeq
where $\bar{u}_c,v_{\tilde{\chi}_1^0}$ denote the charm and neutralino
spinors and $k_1,k_2$ are the four-momenta of the outgoing neutralino and 
charm quark. $F_L$ and $F_R$ are form factors associated with the chirality
projectors ${\cal P}_{L}$ and ${\cal P}_{R}$, respectively. They
receive contributions $F^{v}$ from the vertex diagrams, $F^{tc},
F^{\tilde{t}_1\tilde{c}_L}$  from the quark and squark self-energies,
and $\delta F^{v}, \delta F^{tc}, \delta F^{\tilde{t}_1\tilde{c}_L}$
from the vertex and the quark and squark wave function renormalization
counterterms,
\beq
F_{L,R} = [F^{\tilde{t}_1 \tilde{c}_L } + \delta
F^{\tilde{t}_1 \tilde{c}_L} + F^{tc} + \delta F^{tc} + F^{v} + \delta
F^{v}]_{L,R} \;.
\label{eq:formfac}
\eeq
They are specified in section \ref{sec:renorm} and the Appendices B1-3
and C. 

\section{\label{sec:notation} The quark and squark sector} 
For the choice of our notation the quark and squark sectors are
discussed here in more detail. The definition of the couplings is
deferred to Appendix A.  We define the $3\times 3$ unitary matrices
$U^{u_{L,R}}, U^{d_{L,R}}$ as the matrices which rotate the  left- and
right-handed up- and down-type quark current eigenstates $u_{L,R},
d_{L,R}$ to their corresponding mass eigenstates, $u_{L,R}^m, d_{L,R}^m$, 
\beq
u_L^m = U^{u_L} u_L, \qquad u_R^m = U^{u_R} u_R, \qquad d_L^m = U^{d_L}
d_L, \qquad d_R^m = U^{d_R} d_R \; .
\eeq  
The CKM matrix ${\cal V}$ is given by
\beq
{\cal V} = U^{u_L} U^{d_L\dagger} \,.
\eeq
For the squark interaction eigenstate we define a six component vector 
\beq
\tilde{q}' = \left( \begin{array}{c} \tilde{q}'_L \\
    \tilde{q}'_R \end{array} \right)
\; ,
\eeq
where $\tilde{q}'_L, \tilde{q}'_R$ are each a three component column
vector in generation space.  The squared squark mass matrix can be
written as a $2\times 2$ Hermitian matrix of $3\times 3$ blocks 
\beq
{\cal M}_{\tilde{q}'}^2 = \left( \begin{array}{cc} {\cal
      M}_{\tilde{q}'_{LL}}^2 & {\cal M}_{\tilde{q}'_{LR}}^2 \\
 {\cal M}_{\tilde{q}'_{RL}}^2 &  {\cal M}_{\tilde{q}'_{RR}}^2\end{array}
\right)
\; .
\eeq
It is diagonalized by a $6\times 6$ unitary matrix
$\tilde{W}^q$ which rotates the squark interaction eigenstates to their
mass eigenstates $\tilde{q}^m$, 
\beq
\tilde{q}^m = \tilde{W} \tilde{q}' \; .
\label{eq:fprime}
\eeq
The six component column vector $\tilde{q}^m$ is defined to be ordered  
in mass, with $\tilde{q}_1$ being the lightest
squark. Equation~(\ref{eq:fprime}) can then be rewritten as 
\beq
\tilde{q}^m_s &=& \tilde{W}_{si} \, \tilde{q}'_{iL} + \tilde{W}_{s \,
  i+3} \, \tilde{q}'_{iR}   \qquad \qquad (s=1,..,6,\; i=1,2,3)
\nonumber \\
&\equiv& (\tilde{W}_L \, \tilde{q}'_{L} + \tilde{W}_R \, \tilde{q}'_{R})_s  \;,
\label{eq:sumw}
\eeq
where $i$ denotes the generation
index.\footnote{We have thus decomposed the mass eigenstate squark 
  field into left- and right-chiral interaction eigenstate squarks.}
The rotation of the squarks by the same unitary matrices $U^{q_{L,R}}$
as the quarks defines the super-CKM basis. In models with non-minimal
flavour violation the squark mass matrix is flavour-mixed in this
basis, in contrast to the quark mass matrix. In models with MFV at the
scale $\mu_{\mbox{\scriptsize MFV}}$, however, the squarks can be
rotated by $U^{q_{L,R}}$ to their flavour 
eigenstates in parallel to the quarks, and the super-CKM basis is at
the same time the flavour eigenstate basis. Hence, suppressing
generation indices, 
\beq
\tilde{q}_L = U^{q_L} \tilde{q}'_L, \qquad \tilde{q}_R = U^{q_R}
\tilde{q}'_R \;.
\label{eq:fpf}
\eeq
The squared mass matrix in the flavour eigenstate basis
$(\tilde{q}_L,\tilde{q}_R)^T$ then reads
\beq
{\cal M}^2_{\tilde{q}} = \left( \begin{array}{cc}  (\tilde{M}_{\tilde{q}_L}^2 +
    m_q^2) {\bf 1}_3 &
m_q (A_q - \mu r_q) {\bf 1}_3 \\
m_q (A_q - \mu r_q) {\bf 1}_3 & (\tilde{M}_{\tilde{q}_R}^2 +
m_q^2) {\bf 1}_3 
\end{array} \right) \; ,
\eeq 
where $r_d= 1/r_u=\tan\beta$ for down- and up-type squarks. With
$\tan\beta$ we denote the ratio of the vacuum expectation values of the two
complex Higgs doublets. They are introduced in order to generate masses
of up- and down-type fermions \cite{fayet}. The parameter $A_q$
denotes the trilinear coupling of the soft SUSY breaking part of the 
Lagrangian, $\mu$ the Higgsino mass parameter and $m_q$ the quark
partner mass. ${\bf 1}_3$ is a $3\times 3$ unit matrix in generation
space. The parameters $\tilde{M}_{\tilde{q}_{L,R}}$ are given by the left- and
right-handed scalar soft SUSY breaking masses $M_{\tilde{q}_{L,R}}$ and
the $D$-terms, 
\beq
\tilde{M}^2_{\tilde{q}_{L,R}} &=& M^2_{\tilde{q}_{L,R}} +
D_{\tilde{q}_{L,R}} \nonumber \\
D_{\tilde{q}_{L}} &=&  M_Z^2 \cos 2\beta (I_q^3 - Q_q \sin^2 \theta_W)
\nonumber \\
D_{\tilde{q}_{R}} &=&  M_Z^2 \cos 2\beta \, Q_q \sin^2 \theta_W \;,
\eeq
where $I_q^3$ denotes the third component of the weak isospin, $Q_q$
the electric charge, $M_Z$ the $Z$ boson mass and $\theta_W$ the
Weinberg angle. 
The squared mass matrix ${\cal M}^2_{\tilde{q}}$ can be diagonalized 
by a $6\times 6$ unitary matrix $W^q$ which rotates the flavour
eigenstates to their mass eigenstates,  
\beq
\tilde{q}^m_s &=& W_{st} \left( \begin{array}{c} \tilde{q}_L \\
    \tilde{q}_R \end{array} \right)_t = W_{si} \,\tilde{q}_{Li} +
W_{s\, i+3} \,\tilde{q}_{Ri} \equiv (W_L \,\tilde{q}_{L} +
W_R \,\tilde{q}_{R})_s \label{eq:sfmass} \\
&& \qquad (s,t=1,..,6, \; i=1,2,3) \;. \nonumber
\eeq
Comparison with Eqs.~(\ref{eq:sumw}) and (\ref{eq:fpf}) shows that we 
can factorize the $6\times 3$ matrices $\tilde{W}_{L,R}$ 
into the $6\times 3$ flavour-diagonal matrices $W_{L,R}$ and the $3\times 3$
quark rotation matrices defined above, 
\beq
\tilde{W}_L = W_L U^{q_L}\qquad \mbox{and} \qquad
\tilde{W}_R = W_R U^{q_R} \;,
\label{eq:factor}
\eeq
with $q=u,d$. The matrix $W$ can be expressed in terms of mixing
angles by 
\beq
(W_L)_{ii} = (W_R)_{i+3\, i} = \cos\theta_{q_i} \qquad,\qquad
(W_R)_{ii} = - (W_L)_{i+3\, i} = \sin\theta_{q_i}  \;. 
\label{eq:wmatel}
\eeq
For the three quark generations $i$ the relation between the
flavour eigenstates $\tilde{q}_{iL},\tilde{q}_{iR}$ and the squark mass eigenstates
$\tilde{q}_s^m=(\tilde{q}^m_i,\tilde{q}^m_{i+3})$ hence reads 
\beq
\tilde{q}^m_i &=& \tilde{q}_{iL} \cos\theta_{q_i} + \tilde{q}_{iR} \sin\theta_{q_i}
\nonumber \\
\tilde{q}^m_{i+3} &=& -\tilde{q}_{iL} \sin\theta_{q_i} + \tilde{q}_{iR}
\cos\theta_{q_i}  \; .
\eeq
For better legibility, we suppress the generation indices from now on 
wherever possible, and the lighter and heavier squark mass eigenstates
are generically called $\tilde{q}_1$ and $\tilde{q}_2$. The mixing
angles are then given by 
\beq
\sin 2\theta_q = \frac{2m_q (A_q - \mu r_q)}{M_{\tilde{q}_1}^2 -
  M_{\tilde{q}_2}^2} \qquad , \qquad
\cos 2\theta_q = \frac{\tilde{M}_{\tilde{q}_L}^2 - \tilde{M}_{\tilde{q}_R}^2}
{M_{\tilde{q}_1}^2 - M_{\tilde{q}_2}^2} \; ,
\eeq
and the masses of the squark mass eigenstates read
\beq
M_{\tilde{q}_{1,2}}^2 = m_q^2 + \frac{1}{2} \left[\tilde{M}_{\tilde{q}_L}^2 +
  \tilde{M}_{\tilde{q}_R}^2 \mp \sqrt{(\tilde{M}_{\tilde{q}_L}^2
    -\tilde{M}_{\tilde{q}_R}^2)^2 + 4m_q^2 (A_q-\mu r_q)^2 }  \right] \;.
\eeq
Since the mixing angles are proportional to the masses of the quarks,
the mixing is important in the stop sector and can drive the lightest
stop mass even lighter than the top quark mass. 

\section{\label{sec:renorm} Counterterms and Renormalization} 
The quark and squark self-energies and the vertex diagrams are ultraviolet (UV)
divergent and need to be renormalized. The quark and squark wave
functions are renormalized on-shell. With the bare (s)quark fields 
$q^{(0)}$ ($\tilde{q}^{(0)}$) related to the renormalized (s)quark
fields $q$ ($\tilde{q}$)
by 
\beq
\tilde{q}^{(0)}  = 
\left( 1+\frac{1}{2} \delta Z^{\tilde{q}} \right) \tilde{q} \qquad \mbox{and}
\qquad
q_{L,R}^{(0)} =  \left( 1+\frac{1}{2} \delta Z^{L,R} \right) q_{L,R} \;,
\eeq
this leads for the squarks to the following off-diagonal elements of
the wave function renormalization constants in terms of the real part
of the squark self-energy $\tilde{\Sigma}$,
\beq
\delta Z^{\tilde{q}}_{st} = \frac{2}{m_{\tilde{q}_s}^2-m_{\tilde{q}_t}^2}
\, \mbox{Re} \tilde{\Sigma}_{st} (m_{\tilde{q}_t}^2)  \qquad
s,t=1,..,6, \; s\ne t \;.
\eeq
Defining the following structure for the quark self-energy,
\beq
\Sigma_{ij} (p^2) \equiv \slash{\!\!\! p} \Sigma^L_{ij} (p^2) {\cal P}_L + 
\slash{\!\!\! p} \Sigma^R_{ij} (p^2) {\cal P}_R + m_i 
  \Sigma^{Ls}_{ij} (p^2) {\cal P}_L + m_j \Sigma^{Rs}_{ij} (p^2) {\cal P}_R \; ,
\label{eq:strucself}
\eeq
we have for the corrections to the off-diagonal chiral components of
the quark wave functions, 
\beq
\delta Z^L_{ij} &=& \frac{2}{m_{q_i}^2-m_{q_j}^2} \left[ m_{q_i}^2 \,
  \mbox{Re} \Sigma_{ij}^{Ls} (m_{q_j}^2) + m_{q_j}^2 \,
  \mbox{Re} \Sigma_{ij}^{Rs} (m_{q_j}^2) + m_{q_j}^2 \,
  \mbox{Re} \Sigma_{ij}^{L} (m_{q_j}^2) + m_{q_i} m_{q_j} \,
  \mbox{Re} \Sigma_{ij}^{R} (m_{q_j}^2) \right] \nonumber \\
\delta Z^R_{ij} &=& \frac{2}{m_{q_i}^2-m_{q_j}^2} \left[ m_{q_i} m_{q_j} \,
  \mbox{Re} \Sigma_{ij}^{Ls} (m_{q_j}^2) + m_{q_i} m_{q_j} \,
  \mbox{Re} \Sigma_{ij}^{Rs} (m_{q_j}^2) + m_{q_i} m_{q_j} \,
  \mbox{Re} \Sigma_{ij}^{L} (m_{q_j}^2) + \nonumber \right.\\
& &  \left. m_{q_j}^2 \,
  \mbox{Re} \Sigma_{ij}^{R} (m_{q_j}^2) \right] \qquad\qquad\qquad
 i,j=1,2,3, \; i\ne j \; .
\eeq
As for our scenario we chose the $\tilde{t}_1$ to be the NLSP, all our
self-energies are real, and after the on-shell renormalization of the
quark and squark wave functions we are only left with the one-loop
vertex diagrams and the FCNC vertex counterterm. From now on
``$\mbox{Re}$'' will be dropped. We regularize the divergences by dimensional
regularization in $n=4-2\epsilon$ dimensions so that ultraviolet
singularities appear as poles in $\epsilon$. We have explicitly
verified that the same result is obtained with dimensional reduction.
By exploiting the unitarity relations of the CKM matrix
as well as those of the chargino mixing matrices $U$ and $V$, defined
in Appendix A, we find for the vertex contribution to the form factors
\beq
g \, F^{v}_R &=& - i g\, {\cal F} \left[ 
\frac{1}{\epsilon}  +  \ln \frac{\bar{\mu}^2}{m_{\mbox{\scriptsize
        loop}}^2} + \mbox{finite terms} \right] \label{eq:vertexcontr} \\
g \, F^{v}_L &=& 0 \;,
\eeq
where $\bar{\mu}$ denotes the 't Hooft mass of dimensional
regularization. We have used the short-hand notation 
\beq
{\cal F} \equiv \frac{1}{16 \pi^2} g^2 \sqrt{2}
\left[ \frac{Z_{11}}{6} t_W + \frac{Z_{12}}{2} \right] 
\left( \frac{{\cal V}_{cb}
\, {\cal V}_{tb}^* \, m_b^2 \, \cos \theta_t }{2 M_W^2 c_\beta^2}
\right) \;,
\label{eq:fshort}
\eeq
where $Z_{11},Z_{12}$ denote matrix elements of the $4\times 4$ $Z$
matrix, which diagonalizes the neutralino mass matrix, {\it cf.} Appendix A.
The further 'finite terms' in Eq.~(\ref{eq:vertexcontr}), which do not
depend on $\ln \bar{\mu}^2$, can be extracted from the full one-loop
result for the vertex which is given in Appendix B3. We have 
introduced a generic mass for the loop particles, $m_{\mbox{\scriptsize
    loop}}$. Note, that in the numerical analysis we will use the
exact results with the different loop particle masses. Here, for
reasons of legibility and also to make later contact with the result
derived in Ref.~\cite{Hikasa:1987db} we adopt the generic notation. The
left-handed form factor $F_L$ is zero due to our choice of vanishing
$c$-quark mass. \s

The FCNC counterterm arises from the flavour non-diagonal part of the
wave function renormalization, from the renormalization of the quark and
squark mixing matrices 
\cite{Sirlin:1974ni,Denner:1990yz,Gambino:1998ec,Yamada:2001px} 
and the renormalization of the quark masses.\footnote{In our
  renormalization procedure and the definition of the counterterm we
  follow the same approach as in Ref.~\cite{Degrassi:2006eh}.} 
The renormalized squark
mixing matrix $\tilde{W}^r$ is related to the bare $\tilde{W}^{(0)}$
by 
\beq
\tilde{W}^{(0)}_{su} = (\delta_{st} + \delta\tilde{w}_{st})
\tilde{W}^{r}_{tu} \;, \qquad s,t,u=1,..,6.
\eeq
And similarly, the renormalized quark mixing matrices $U^{u_{L,R} \, r}$
are related to the bare quark matrices $U^{u_{L,R}\, (0)}$ by 
\beq
U^{u_{L,R}\, (0)}_{ik} = (\delta_{ij}+ \delta u^{u_{L,R}}_{ij}) U^{u_{L,R}\, r}_{jk}
\;, \qquad i,j,k=1,2,3. 
\eeq
The indices $s,t,u$ denote the six squark mass eigenstates, and $i,j,k$ are
generation indices. We impose the MFV condition on the renormalized
mixing matrices $\tilde{W}^r,U^r$ and hence demand them to be flavour
diagonal. This leads to flavour 
non-diagonal counterterms $\delta \tilde{w}, \delta
u^{u_{L,R}}$. Furthermore, as the bare and renormalized mixing matrices 
are unitary the counterterms must be antihermitian. The UV divergent
part of each counterterm is determined such that it cancels the
divergent part of the antihermitian part of the corresponding wave
function renormalization matrix 
\cite{Denner:1990yz,Gambino:1998ec,Yamada:2001px},
\beq
\delta \tilde{w} &=& \frac{1}{4} (\delta Z^{\tilde{q}} -
\delta Z^{\tilde{q} \dagger}) 
\label{eq:wcounter}
\\
\delta u^{u_{L,R}} &=& \frac{1}{4} (\delta Z^{L,R} - \delta
Z^{L,R \dagger}) \;.
\label{eq:ucounter}
\eeq
The general form of the FCNC vertex counterterm depicted in
Fig.~\ref{fig:counter} has been derived in Appendix C. For our process we 
have the following form factor contributions to the vertex counterterm,
\beq
g \, \delta F^v_{R} &=& 
-ig \, e^u_{L1} \left[ \frac{1}{2} \delta Z^{L\dagger}_{ct}
  \cos\theta_t + \frac{1}{2} 
\delta Z^{\tilde{q}}_{\tilde{c}_L \tilde{t}_1} + \delta u^{U_L}_{ct}
\cos\theta_t + \delta\tilde{w}^\dagger_{\tilde{c}_L
  \tilde{t}_1}\right] 
\label{eq:delfv} \\
g \, \delta F^v_{L} &=& 0 \;,
\eeq
with
\beq
e^u_{L1} = \sqrt{2} \left[ \frac{Z_{11}}{6} \tan \theta_W +
  \frac{1}{2} Z_{12} \right]
\eeq
and
\beq
\delta Z^{L\dagger}_{ct} &=& 2 \Sigma^{Ls}_{tc} (m_c^2=0) = 0 
\label{eq:delzl} \\
\delta Z^{\tilde{q}}_{\tilde{c}_L\tilde{t}_1} &=&
\frac{2\Sigma^{\tilde{q}}_{\tilde{c}_L \tilde{t}_1}
  (\mst1^2)}{\msc1^2-\mst1^2}   \;.
\eeq
The squark wave function has been renormalized at $p^2=
m_{\tilde{t}_1}^2$. The finite parts of the counterterms
Eqs.~(\ref{eq:wcounter},\ref{eq:ucounter}) depend on the
renormalization conditions. Absorbing also the finite part of the
antihermitian wave function renormalization leads to a gauge-dependent
on-shell renormalization scheme
\cite{Yamada:2001px,Gambino:1998ec,Kniehl:2000rb}. 
Performing minimal subtraction on the other hand is gauge-independent
\cite{Gross:1973ju} and imposes the MFV condition on the
$\overline{\mbox{MS}}$ parameters at the scale $\mu_{\mbox{\scriptsize
    MFV}}$. In the following we will adopt this scheme. Consequently, the
result will depend on the MFV scale $\mu_{\mbox{\scriptsize
    MFV}}$. The squark mixing matrix counterterm then
reads\footnote{This counterterm definition can lead to large contributions
  to the matrix element if $m_{\tilde{t}_1} \approx
  m_{\tilde{c}_L}$. For a discussion, see {\it e.g.}
  Refs.~\cite{Eberl:1999he}. In the scenarios of our numerical
  analysis we have $m_{\tilde{t}_1} \ll m_{\tilde{c}_L}$.} 
\beq
\delta \tilde{w}^\dagger_{\tilde{c}_L \tilde{t}_1} &=& \frac{1}{2} \left( 
\frac{\Sigma^{\tilde{q}}_{\tilde{c}_L \tilde{t}_1} (\msc1^2) + 
\Sigma^{\tilde{q}}_{\tilde{c}_L \tilde{t}_1}
(\mst1^2)}{\mst1^2-\msc1^2}\right)_{\scriptsize \overline{\mbox{MS}}} \;.
\eeq
A gauge invariant prescription for the quark mixing matrix counterterm
is given by \cite{Gambino:1998ec},
\beq
\delta u^{U_L}_{ct} &=& -\frac{1}{2} \left[ \Sigma^L_{tc} (0)
  +2\Sigma^{Ls}_{tc} (0)  \right]_{\scriptsize \overline{\mbox{MS}}}  
=  -\frac{1}{2} 
  \Sigma^L_{tc} (0)_{\scriptsize \overline{\mbox{MS}}} \;.
\eeq
For the quark mixing matrix contribution to the vertex counterterm we
then find  
\beq
-g \, e^u_{L1} \cos\theta_t \left(\delta u^{U_L}_{ct}
\right)_{\overline{\mbox{\scriptsize{}MS}}} = g\,
\frac{{\cal F}}{2} \left[ \frac{1}{\epsilon} + \ln 
  \frac{\bar{\mu}^2}{\mu_{\mbox{\scriptsize MFV}}^2} \right] \; .
\label{eq:deluL}
\eeq
For the contribution from the squark mixing we have 
\beq
-g \, e^u_{L1} \left(\delta \tilde{w}^\dagger_{\tilde{c}_L \tilde{t}_1}
\right)_{\overline{\mbox{\scriptsize{MS}}}} = g\, \frac{{\cal F}}{2}
\left( \frac{-\mst1^2-\msc1^2-2{\cal A}}{\mst1^2-\msc1^2} \right)
\left[ \frac{1}{\epsilon} +  \ln
  \frac{\bar{\mu}^2}{\mu_{\mbox{\scriptsize MFV}}^2} \right] \; ,
\label{eq:delw}
\eeq
with
\beq
{\cal A} = -\mu^2 + A_b^2 + \tilde{M}_{\tilde{b}_R}^2 + c_\beta^2 (M_W^2
  (t_\beta^2-1) + M_A^2 t_\beta^2) + m_t A_b \tan\theta_t
\;.
\label{eq:ashort}
\eeq
It depends on the Higgsino parameter $\mu$, the soft SUSY breaking
mass parameter $\tilde{M}_{\tilde{b}_R}$ including $D$ term
contributions, the trilinear coupling $A_b$, the 
mixing angle $\beta$, the $W$ boson mass and the pseudoscalar Higgs mass
$M_A$. \s

The contribution from $\delta Z^{\tilde{q}}_{\tilde{c}_L\tilde{t}_1}$ is given by
\beq
-g \, e^u_{L1}  \frac{1}{2} \left(  \delta  Z^{\tilde{q}}_{\tilde{c}_L\tilde{t}_1}
\right) =
g\, 
\frac{{\cal F}}{2} \left( \frac{2\mst1^2+2 {\cal A}}{\mst1^2-\msc1^2} \right)
\left[ \frac{1}{\epsilon} + \ln \frac{\bar{\mu}^2}{m_{\mbox{\scriptsize
        loop}}^2} + \mbox{finite terms} \right]  \;.
\label{eq:delztilde}
\eeq
As before, we have introduced a generic loop particle mass, and 'finite
terms' denote further terms which do not depend on $ \ln
\bar{\mu}^2$. Their specific form can be extracted from the explicit
formulae of the stop self-energies given in Appendix B1. Inserting
Eqs.~(\ref{eq:delzl},\ref{eq:deluL},\ref{eq:delw},\ref{eq:delztilde})
in Eq.~(\ref{eq:delfv}), the right-chiral part of the FCNC counterterm is
then given by
\beq
g \, \delta F^v_{R} = i g\, {\cal F}
\left[ \frac{1}{\epsilon} - \frac{\msc1^2+{\cal A}}{\mst1^2-\msc1^2}
  \ln \frac{\bar{\mu}^2}{\mu_{\mbox{\scriptsize MFV}}^2} + \frac{\mst1^2+{\cal
    A}}{\mst1^2-\msc1^2} \ln \frac{\bar{\mu}^2}{m_{\mbox{\scriptsize
        loop}}^2} + \mbox{finite terms} \right] \;.
\label{eq:deltavcontr}
\eeq
Adding Eqs.~(\ref{eq:vertexcontr}) and (\ref{eq:deltavcontr}) and
replacing ${\cal F}$ and ${\cal A}$ by Eq.~(\ref{eq:fshort}) and
Eq.~(\ref{eq:ashort}), respectively, we arrive at the following final
result for the form factors, which contribute to Eq.~(\ref{eq:formfac}), 
\beq
g\, F_R &=& \frac{i}{16 \pi^2} g^3 \sqrt{2} \left[ \frac{Z_{11}}{6} t_W +
  \frac{Z_{12}}{2} \right] \left( \frac{{\cal V}_{cb} \, {\cal
      V}_{tb}^* \, m_b^2 \, \cos\theta_t }{2 M_W^2 c_\beta^2} \right)
\left(\frac{1}{\mst1^2-\msc1^2} \right) \times
\nonumber \\
&&
[\msc1^2 -\mu^2 + A_b^2 + \tilde{M}_{\tilde{b}_R}^2 + c_\beta^2 (M_W^2
(t_\beta^2-1) + M_A^2 t_\beta^2) + m_t A_b \tan \theta_t ]
\ln \left(\frac{\mu_{\mbox{\scriptsize MFV}}^2}{m_{\mbox{\scriptsize
        loop}}^2}\right) 
\label{eq:formfacresr} \\
&& + \mbox{finite terms} \nonumber
\\
g\, F_L &=& 0 \;.
\label{eq:formfacresl}
\eeq
Finally, the stop decay width in terms of the form factors
Eqs.~(\ref{eq:formfacresr},\ref{eq:formfacresl}) is given by
\beq
\Gamma (\tilde{t}_1\to c \tilde{\chi}_1^0) =
\frac{g^2\mst1}{16 \pi} \left( 1 -\frac{m_{\tilde{\chi}_1^0}^2}{\mst1^2}
\right)^2 |F_R|^2 \;.
\eeq
As can be inferred from $F_R$, depending on the scale of MFV,
the logarithm can become very large, and the decay can become
important in certain regions of the parameter space, especially for
large values of $\tan\beta$. The finite terms, which do not depend on
$\ln \mu_{\mbox{\scriptsize MFV}}^2$, are then only
subleading. If we drop the finite terms in Eq.~(\ref{eq:formfacresr}), 
the approximate result given by Hikasa and Kobayashi 
in Ref.~\cite{Hikasa:1987db} should be
reproduced. In fact, for $m_c=0$ we can rewrite
\beq
\msc1^2 -\mu^2 + c_\beta^2 (M_W^2 (t_\beta^2-1) + M_A^2 t_\beta^2) 
= M_{H_d}^2 + M_{\tilde{q}_L}^2 + \frac{1}{3} M_Z^2 \sin^2 \theta_W \cos
2\beta \;,
\label{eq:relation}
\eeq
where $M_{H_d}$ denotes the mass parameter of the Higgs doublet $H_d$
which couples to down-type fermions. With this relation
the form factor Eq.~(\ref{eq:formfacresr}) leads to the
approximate result $F_R^{H/K}$ of Ref.~\cite{Hikasa:1987db}, if we set
the MFV scale equal to the Planck scale, $\mu_{\mbox{\scriptsize MFV}}
= M_P$, choose $M_W$ as generic loop particle mass and neglect the
finite terms, 
\beq
g F_R^{H/K} &=& \frac{i}{16 \pi^2} g^3 \sqrt{2} \left[ \frac{Z_{11}}{6} t_W +
  \frac{Z_{12}}{2} \right] \left( \frac{{\cal V}_{cb} \, {\cal
      V}_{tb}^* \, m_b^2 \, \cos\theta_t }{2 M_W^2 c_\beta^2} \right)
\left(\frac{1}{\mst1^2-\msc1^2} \right) \times
\nonumber \\
&&
[M_{H_d}^2 + M_{\tilde{q}_L}^2 + A_b^2 + M_{\tilde{b}_R}^2 + m_t A_b
\tan \theta_t ] \ln \left(\frac{M_P^2}{M_W^2}
\right) \label{eq:hkformfacresr} 
\;.
\eeq 
In our full one-loop calculation of the decay width in terms
of $F_R$, Eq.~(\ref{eq:formfacresr}), the finite terms are included,
and the relevance of these contributions can  
be checked by comparing with the approximate result $\Gamma^{H/K}$ for
the decay width in terms of the form factor $F_R^{H/K}$,
Eq.~(\ref{eq:hkformfacresr}). This will be discussed in section
\ref{sec:numerical}. \s  

To get a reliable result, the large logarithms of the MFV scale in the
decay formula should be resummed. The logarithm is related to the
running of the FCNC coupling of the neutralino to a quark and squark
of different generations. We have required this coupling to vanish at the scale
$\mu_{\mbox{\scriptsize{MFV}}}$. Minimal Flavour Violation is not
RGE-invariant, however. Even though MFV is imposed at
$\mu_{\mbox{\scriptsize{MFV}}}$,  at any other scale $\mu \ne
\mu_{\mbox{\scriptsize{MFV}}}$ a FCNC coupling will be generated
through renormalization group evolution. The solution of the one-loop
RGE for the quark and squark mixing matrices provides the resummation
of the large $\ln \mu^2_{\mbox{\scriptsize{MFV}}}$. The coefficient of
this logarithm in 
Eq.~(\ref{eq:formfacresr}) is then given by the first order in the
expansion of the RGE solution for the squark-quark-neutralino coupling
in powers of $\alpha$. In the following we will call the right-handed
form factor including the resummation effects $F_R^{FV}$. It is
given by the FCNC coupling obtained through renormalization group evolution
including flavour violation of the squark and quark mixing matrices,
from some high scale down to the scale relevant for the decay process. \s

\section{\label{sec:numerical} Numerical Analysis} 
The scenarios for the numerical analysis have been chosen such that
they lead to a NLSP stop $\tilde{t}_1$ and a $\tilde{\chi}_1^0$
LSP. The latter represents a promising dark matter (DM)
candidate in the MSSM \cite{Ellis:1983ew}. The mass difference is
chosen to be small enough so that the loop mediated flavour changing
decay $\tilde{t}_1 \to c \tilde{\chi}_1^0$ is dominating and can
compete with the four-body decays\footnote{Scenarios where 2-body
  decays at tree level are forbidden for the lightest stop quark and where
  the loop induced flavour changing decay competes with 3-body decays have
  been discussed in \cite{Porod:1996at}.}  into the LSP, a $b$-quark and
a fermion pair \cite{Boehm:1999tr}, 
\beq
\tilde{t}_1 \to \tilde{\chi}_1^0 b f \bar{f}' \; .
\label{eq:4body}
\eeq
Such scenarios can be consistent with electroweak baryogenesis
\cite{EWBG} and 
also with Dark Matter constraints \cite{dmconstr}. The mass spectra
and mixing angles have been calculated with the spectrum calculator SPheno
\cite{Porod:2003um} and compared to SOFTSUSY \cite{Allanach:2001kg}. 
Both codes include the option to perform two-loop RGE running with and
without the inclusion of flavour violation and both support the
SUSY Les Houches accord \cite{Skands:2003cj}. Within this accord the
gauge and Yukawa couplings as well as the soft SUSY breaking mass
parameters and trilinear couplings are given out as
$\overline{\mbox{DR}}$ running parameters at a scale $Q$, which we
have chosen to be the scale of electroweak symmetry breaking
(EWSB). We have verified that the calculation 
of the decay width leads to the same result if we apply dimensional
reduction instead of dimensional regularization, so that the
$\overline{\mbox{DR}}$ running parameters can be used. 
The mixing matrix elements and the SUSY particle pole masses
have been taken at the scale of EWSB as well. The SM parameters have
been chosen as $M_Z = 91.187$ GeV,
$\alpha_{em}^{-1\overline{\mbox{\scriptsize MS}}} (M_Z)= 127.934$,
$\alpha_{s}^{\overline{\mbox{\scriptsize MS}}} (M_Z)= 0.1184$,
$m_b^{\overline{\mbox{\scriptsize MS}}} (m_b) = 4.25$ GeV,
$M_t^{\mbox{\scriptsize{pole}}} = 173.3$ GeV and
$m_\tau^{\mbox{\scriptsize{pole}}} = 1.777$ GeV. We have chosen the CKM matrix elements as  $|{\cal V}_{tb}| = 0.9993$ and $|{\cal V}_{cb}| = 0.04$. 
In order to ensure MFV, for all three generations a common mass parameter 
$M_{\tilde{q}_L}$ for the soft SUSY breaking masses of the $SU(2)$
doublet has to be introduced at the scale $\mu=\mu_{\mbox{\scriptsize
    MFV}}$, so that the up- and down-type squark mass matrices can be
simultaneously flavour-diagonal. We work in the framework of a MFV
MSSM defined at the GUT scale in terms of a small number of
parameters. They are given by common soft SUSY breaking scalar
and gaugino mass terms, $M_0$ and $M_{1/2}$, a common SUSY breaking
trilinear coupling $A_0$, the ratio of the two vacuum expectation
values $\tan\beta$ and the sign of the Higgsino parameter 
$\mu$. \s 

\subsection{Analysis for \boldmath{$\mu_{\mbox{\scriptsize MFV}}
    \approx 10^{16}$} GeV}
We first investigate two mSUGRA scenarios with soft-breaking terms
at the GUT scale $M_{GUT} \approx 10^{16}$~GeV, which is identified
with the MFV scale. All soft SUSY breaking parameters are family universal. 
The boundary conditions at $\mu_{\mbox{\scriptsize MFV}} =M_{GUT}$ are 
\beq
\begin{array}{llll}
(1) & M_0 = 200 \mbox{ GeV} & M_{1/2} = 230 \mbox{ GeV} & A_0 = -920
\mbox{ GeV} \\
& \tan\beta = 10 & \mbox{sign}(\mu) = + \\
(2) & M_0 = 200 \mbox{ GeV} & M_{1/2} = 230 \mbox{ GeV} & A_0 = -895
\mbox{ GeV} \\
& \tan\beta = 10 & \mbox{sign}(\mu) = + \;.
\end{array}
\label{eq:scenarios}
\eeq
The second scenario has a larger $\tilde{t}_1-\tilde{\chi}_1^0$ mass
difference compared to scenario (1), whereas the mass difference
between $\tilde{t}_1$ and the lightest chargino $\tilde{\chi}_1^+$ is
smaller. Since the 4-body decays are dominated by the chargino
exchange diagram \cite{Boehm:1999tr}, in 
scenario (2) the 4-body decays should be more important leading to a 
smaller branching ratio of the flavour changing decay. The GUT scale
is given by $\sim 2.3 \cdot 10^{16}$ GeV. The masses are obtained
by RGE evolution from the GUT scale down to the electroweak scale. The
running is performed at two-loop order without the inclusion of
explicit flavour violation in the squark sector. The obtained masses are
\beq
\begin{array}{llll}
(1) & \mst1 = 104 \mbox{ GeV} &
m_{\tilde{\chi}_1^0} = 92 \mbox{ GeV} 
& m_{\tilde{\chi}_1^+} = 175 \mbox{ GeV} \\
(2) & \mst1 = 130 \mbox{ GeV} &
m_{\tilde{\chi}_1^0} = 92 \mbox{ GeV} 
& m_{\tilde{\chi}_1^+} = 175 \mbox{ GeV} \;.
\end{array}
\label{eq:scen12}
\eeq
For these scenarios the partial stop decay width into charm and
neutralino, calculated with the full one-loop formula, is compared to
the approximate result. For the latter, we take $M_W$ as generic loop
particle mass, {\it cf.} Eq.~(\ref{eq:hkformfacresr}). The widths and
form factors are given in Table 1. They have been obtained with the
program SUSY-HIT \cite{susyhit}, where the full one-loop formula for
the flavour changing stop decay has been implemented. 
\begin{table}[ht]
\vspace*{0.3cm}
\begin{center}
$
\renewcommand{\arraystretch}{1.5}
\begin{array}{|c|c|c||c|c|} \hline
\tilde{t}_1 \to c \tilde{\chi}_1^0 & \Gamma^{\mbox{\scriptsize{1-loop}}}
\mbox{[GeV]} & |F_R^{\mbox{\scriptsize{1-loop}}}| & \Gamma^{\mbox{\scriptsize{H/K}}}
\mbox{[GeV]}  & |F_R^{\mbox{\scriptsize H/K}}| \\ \hline\hline
\mbox{Scenario} (1) & 9.322 \cdot 10^{-10} & 1.486 \cdot 10^{-4} & 1.004 \cdot
10^{-9} &  1.542 \cdot 10^{-4} \\ \hline
\mbox{Scenario} (2) & 5.862 \cdot 10^{-9} & 1.460 \cdot 10^{-4} & 6.446 \cdot
10^{-9} & 1.531 \cdot 10^{-4} \\ \hline
\end{array}
$
\caption{\label{table:partwidth} The partial widths and form factors
  for the decay $\tilde{t}_1\to  c \tilde{\chi}_1^0$ in two MFV
  scenarios, calculated with the exact 1-loop formula,
  $\Gamma^{\mbox{\scriptsize{1-loop}}}$, $F_R^{\mbox{\scriptsize{1-loop}}}$,
  and with the approximate formula of Ref.~\cite{Hikasa:1987db},
  $\Gamma^{\mbox{\scriptsize{H/K}}}$,
  $F_R^{\mbox{\scriptsize{H/K}}}$.} 
\end{center}
\vspace*{-0.2cm}
\end{table}
As can be inferred from the table, the exact and approximate decay
width differ by ${\cal O}(10)$\%. In fact, the finite terms 
extracted from the one-loop formula turn out to contribute with $\sim 
3-5$\% to $F_R$, Eq.~(\ref{eq:formfacresr}). This difference leads to
the 10\% effect in the decay width. The difference in the branching ratio
$BR(\tilde{t}_1 \to c \tilde{\chi}_1^0)$ calculated in the two
approaches is negligible, however. We note, that in the first scenario the partial
width is $\sim 6$ times smaller than in the second scenario due the smaller
$\tilde{t}_1-\tilde{\chi}_1^0$ mass difference and hence reduced phase
space. \s 

For the calculation of the branching ratios, also the partial
width for the $\tilde{t}_1$ decay into $u$-quark and neutralino,
$\tilde{t}_1 \to u \tilde{\chi}_1^0$, as well as the 4-body decay width
are needed.  The former is suppressed by 2 orders of magnitude compared
to the $c \tilde{\chi}_1^0$ final state due to the small CKM matrix
element $|{\cal V}_{ub}| \approx 0.003$ which enters quadratically in
the decay width. The branching ratios are listed in Table 2. As
anticipated, the stop 4-body decay is more important in scenario (2)
leading to a change of the branching ratio of interest,
$\mbox{BR}(\tilde{t}_1 \to \tilde{\chi}_1^0 c)$, at the few per-cent level. \s
\begin{table}[h]
\vspace*{0.3cm}
\begin{center}
$
\renewcommand{\arraystretch}{1.5}
\begin{array}{|c|c|c|c|} \hline
\mbox{branching ratio} & \mbox{BR}(\tilde{t}_1 \to \tilde{\chi}_1^0 c)
& \mbox{BR}(\tilde{t}_1 \to \tilde{\chi}_1^0 u)
& \mbox{BR}(\tilde{t}_1 \to \tilde{\chi}_1^0 b f \bar{f}') \\ \hline\hline
\mbox{Scenario} (1) & 0.9944  & 0.0056 & 4.587 \cdot 10^{-5}\\ \hline
\mbox{Scenario} (2) & 0.9443  & 0.0053 & 0.0504 \\ \hline
\end{array}
$
\caption{\label{table:br} The $\tilde{t}_1$ branching ratios for
  different final states for scenario (1) and (2).}
\end{center}
\vspace*{-0.2cm}
\end{table}

As stated before, the large logarithms in the decay formula should be
resummed. To get an estimate of the importance of the resummation
effects, the one-loop decay calculated in the framework of MFV is
compared to the tree level stop decay into charm and neutralino with flavour
off-diagonal elements in the squark mixing matrix. They are the result
of the small flavour off-diagonal entries introduced in the soft-breaking terms
through RG evolution including the complete flavour structure of the
different flavour matrices,  from the scale of MFV down to the scale of EWSB. 
The input parameters for the decay formula
are taken from SPheno \cite{Porod:2003um}\footnote{Thanks to Werner
  Porod who provided us with the newest SPheno version 3.0.beta56.}.  
In the flavour violating case, denoted by FV in the following, where no
flavour-eigenstates exist any more, the lightest up-type squark state
$\tilde{u}_1$ has been identified to correspond to
$\tilde{t}_1$. The scenarios in Eq.~(\ref{eq:scenarios}) have been chosen
such that $m_{\tilde{u}_1} \approx m_{\tilde{t}_1}$. The
$\tilde{\chi}_1^0$ and $\tilde{\chi}_1^+$ masses are almost unchanged.
The form factor $F_R^{\mbox{\scriptsize FV}}$ of the tree level
decay is given by the right-handed part of the FCNC
$\tilde{u}_1-c-\tilde{\chi}_1^0$ coupling\footnote{The left-handed part
  is negligibly small for $m_c=0$.}, 
\beq
F_R^{\mbox{\scriptsize{FV}}} = - i\,\sqrt{2} \left(\frac{Z_{11}}{6}
             t_W + \frac{Z_{12}}{2} \right) (\tilde{W}_L)_{\tilde{u}_1
             c} \;,
\label{eq:frtree}
\eeq
with the squark mixing matrix $\tilde{W}_L$ defined in
Eq.~(\ref{eq:sumw}). This leads to the partial decay width
\beq
\Gamma^{\mbox{\scriptsize{FV}}} (\tilde{u}_1\to c \tilde{\chi}_1^0) =
\frac{g^2 m_{\tilde{u}_1}}{16 \pi} \left( 1 -
  \frac{m_{\tilde{\chi}_1^0}^2}{m_{\tilde{u}_1}^2} \right)
|F_R^{\mbox{\scriptsize{FV}}}|^2 \;. 
\eeq
The form factors and partial widths are shown in Table~\ref{table:compfvmfv}.
As can be inferred from the table, there is a factor $\sim 4.4$ between the
right-handed form factor calculated at the one-loop level in the MFV
framework and the one derived from RG evolution including flavour
\begin{table}
\vspace*{0.3cm}
\begin{center}
$
\renewcommand{\arraystretch}{1.5}
\begin{array}{|c|c|c|c|c|} \hline
 & |F_R^{\mbox{\scriptsize{1-loop}}}| & |F_R^{\mbox{\scriptsize{FV}}}| &
 \Gamma^{\mbox{\scriptsize{1-loop}}} \mbox{ [GeV]} & \Gamma^{\mbox{\scriptsize{FV}}}
 \mbox{ [GeV]}\\ \hline\hline
\mbox{Scenario (1)} & 1.486 \cdot 10^{-4} &  3.361 \cdot 10^{-5} 
& 9.322 \cdot 10^{-10} & 4.766 \cdot 10^{-11} \\ \hline 
\mbox{Scenario (2)} & 1.460 \cdot 10^{-4} & 3.306 \cdot 10^{-5} & 5.862 \cdot
10^{-9} & 3.006 \cdot 10^{-10} \\ \hline 
\end{array}
$
\caption{\label{table:compfvmfv} The right-handed form factors and
  partial decay widths of the lightest up-type squark into charm and
  neutralino for the MFV scenario (1-loop) and the FCNC tree level decay (FV).}
\end{center}
\vspace*{-0.2cm}
\end{table}
violation. As expected, resummation effects turn out to be important for a
large scale $\mu_{\mbox{\scriptsize{MFV}}}=M_{\mbox{\scriptsize
    GUT}}$\footnote{This result is in agreement with the discussion in
  Ref.~\cite{Lunghi:2006uf} where resummation effects in the coupling
  $\tilde{t}_1-c-\tilde{\chi}_1^0$ have been found to be large.}. 
The partial widths, which depend quadratically on the  right-handed
form factor, differ by a factor $\sim 20$. \s

For comparison we have performed the calculation with the decay
spectra and mixing angles evaluated by SOFTSUSY. The squark mixing
matrix elements agree within $10^{-2}$ accuracy with the results of
SPheno. The mixing matrix element $(\tilde{W}_L)_{ \tilde{u}_1 c}$,
which enters in the form factor $F_R^{\mbox{\scriptsize{FV}}}$
Eq.~(\ref{eq:frtree}), is ${\cal O}(10^{-4})$ and differs in the two
spectrum calculators. The two codes implement the one-loop corrections
to the squark mass matrices differently. SOFTSUSY corrects only the
flavour-diagonal entries of the squark mass matrices, while SPheno
implements a full one-loop calculation, so that differences in the
flavour off-diagonal entries are to be expected. For the SOFTSUSY
parameter values, this results in a ratio between loop decay and tree
level decay of $\sim 2.7$ for the two scenarios, compared to the
ratio $\sim 4.4$ found with the SPheno parameter values. All in all,
the results with both spectrum calculators show the importance of the
resummation effects. \s 

Of phenomenological interest are the consequences of these resummation
effects on the $\tilde{t}_1$ branching ratio into the charm plus
neutralino final state. To quantify this, the competing 4-body stop
decay width, calculated in the FV scenario and including tree level
FCNC couplings, is needed. The additional FCNC contributions are
expected to be small, however, due to the suppression by CKM matrix
elements. As the calculation is not available at present we defer a
detailed comparison to a future publication. 
 
\subsection{Analysis for \boldmath{$\mu_{\mbox{\scriptsize      MFV}} \le M_{\mbox{\scriptsize GUT}}$}}
In the previous section the importance of resummation effects has been
discussed. With decreasing $\mu_{\mbox{\scriptsize    MFV}}$ and hence smaller $\ln
\mu_{\mbox{\scriptsize MFV}}^2$ on the other hand the non-resummed
one-loop MFV result should approach the resummed flavour-violating
tree level result. Furthermore, we expect the approximate formula
of Ref. \cite{Hikasa:1987db}, which is a good approximation of the exact
one-loop MFV result for large scales, to be less good with decreasing
MFV scale.  In order to verify this behaviour we have chosen
scenarios with different $\mu_{\mbox{\scriptsize MFV}}$ varied between
$10^3 \mbox{ GeV} \le \mu_{\mbox{\scriptsize MFV}} \le 10^{16}$
GeV. We choose the soft SUSY breaking input parameters\footnote{In our
scenarios $\tan\beta$ varies between 10 and 20.} in each scenario such
that the masses for $\tilde{t}_1$ and $\tilde{\chi}_1^0$ remain almost
unchanged. Consequently, the differences in the partial decay
widths will not be due to phase space effects. Furthermore, the
scenarios are constrained by the requirement that $\tilde{t}_1$ is the
NLSP and $\tilde{\chi}_1^0$ is the LSP, so that the FC decay
$\tilde{t}_1 \to c \chi_1^0$ dominates. The relevant particle masses for
the different scenarios vary as 
\beq
\mst1 = 105 \, ... \, 116 \, \mbox{GeV} \quad \mbox{ and } \quad \mn1 = 92
\, ... \, 104 \, \mbox{GeV} \;,
\eeq 
with the $\tilde{t}_1-\tilde{\chi}_1^0$ mass difference ranging between
\beq
\mst1 - \mn1 = 9 \, ... \, 15 \, \mbox{GeV} \;.
\eeq
We emphasize that the following results are purely
illustrative. The various scenarios have not been required to fulfill
Dark Matter constraints and/or constraints from electroweak precision
data. The 
main emphasis was to achieve approximately constant masses for the
NLSP and LSP. \s

\begin{figure}[t]
\begin{center}
\epsfig{figure=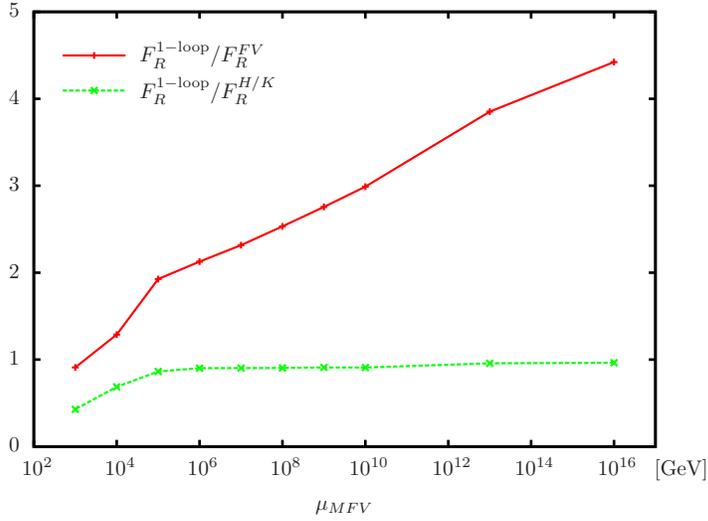,width=9.3cm,clip=}
\caption{\label{fig:stopfac} Ratio between the right-handed form factor of
  the MFV loop decay $F_R^{1-loop}$ and the form factor of the FV tree
  level decay $F_R^{FV}$ (red/full) and ratio between the MFV loop-decay form
  factor and the  approximate form factor $F_R^{H/K}$ 
  (green/dashed) as function of the MFV scale $\mu_{\mbox{\scriptsize MFV}}$.}
\end{center}
\end{figure}

In Fig.~\ref{fig:stopfac} we show, as a function of the MFV scale, the
ratio of the non-resummed right-handed form factor
$F_R^{\mbox{\scriptsize 1-loop}}$ in the MFV 1-loop decay to
$F_R^{\mbox{\scriptsize FV}}$ in the FV tree level decay as well as
the ratio of $F_R^{\mbox{\scriptsize 1-loop}}$ to the approximate form
factor $F_R^{\mbox{\scriptsize H/K}}$.\footnote{Note that the line
  connecting the different points uniquely serves to guide the eye.}  
As can be inferred from the
figure, the approximate result reproduces the one-loop result down to
low scales. Starting from $\mu_{\mbox{\scriptsize MFV}} = 10^5$ GeV
the finite terms become relevant. At $\mu_{\mbox{\scriptsize MFV}} =
10^3$ GeV neglecting the finite terms in 
$F_R^{\mbox{\scriptsize H/K}}$ leads to a factor $\sim 2$ between the
approximate and the 1-loop form factor. The
non-resummed 1-loop result and the resummed tree level result, on the
other hand, approach each other with decreasing scale of MFV as
expected. \s 

\begin{figure}[t]
\begin{center}
\hspace*{-0.6cm}
\epsfig{figure=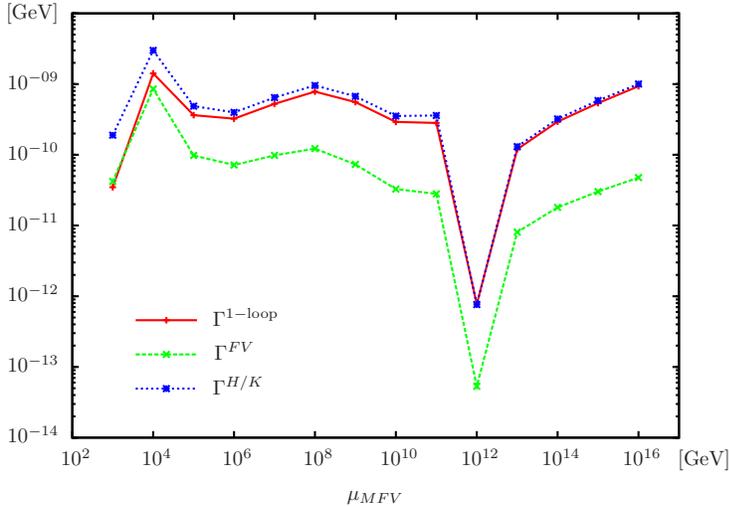,width=9.6cm,clip=}
\caption{\label{fig:stopamp} Partial decay width $\Gamma(\tilde{t}_1\to
  c \tilde{\chi}_1^0)$ calculated assuming MFV, $\Gamma^{1-loop}$
  (red/full), calculated with the approximate formula, $\Gamma^{H/K}$
  (blue/dotted), and calculated at tree level including FV,
  $\Gamma^{FV}$ (green/dashed), as function of $\mu_{\mbox{\scriptsize MFV}}$.}
\end{center}
\end{figure}
Figure \ref{fig:stopamp} shows the partial widths as functions of
$\mu_{\mbox{\scriptsize MFV}}$ for the approximate MFV decay, for the
full MFV 1-loop decay and for the tree level resummed decay. 
An interesting feature which can be inferred from
Fig.~\ref{fig:stopamp} is the size of the decay width. It does not
only depend on the size of the logarithm but also on the coefficient
of the logarithmic term, which is given in terms of the soft SUSY
breaking parameters, particle masses and mixing angles, {\it cf.}
Eq.~(\ref{eq:formfacresr}). As explained above, for each value of the
scale $\mu_{\mbox{\scriptsize MFV}}$ we have chosen a different set of
boundary conditions $M_0,m_{1/2},A_0,\tan\beta,\mbox{sign}(\mu)$ such
that the $\tilde{t}_1$ and $\tilde{\chi}_1^0$ masses remain
approximately unchanged. This leads for each $\mu_{\mbox{\scriptsize
    MFV}}$ to a different coefficient of the logarithmic term. For
$\mu_{\mbox{\scriptsize    MFV}}=10^{12}$ GeV {\it e.g.} the parameter
set and resulting masses and mixing angles are such that the
coefficient becomes rather small, so that the partial width is less
than $10^{-12}$ GeV. Due to the large value of $\mu_{\mbox{\scriptsize
    MFV}}$ the logarithmic contribution still dominates over the
finite terms, however, so that there is good agreement between the 1-loop and
approximate result. For small values of $\mu_{\mbox{\scriptsize
    MFV}}$ the partial width can be as large as a few $10^{-11}$ GeV as
the factor, which multiplies the logarithm, turns out to be large for
the chosen parameter set. The value of the coefficient is also the reason for
the kink in Fig.~\ref{fig:stopfac} at $\mu_{\mbox{\scriptsize
    MFV}}=10^5$ GeV. \s

Figure \ref{fig:stopamp} shows, that in
accordance with the behaviour of the right-handed form factors, at
high scales the 1-loop and the approximate result agree up to the
effect of the non-logarithmic terms on the partial width, which is at the
10\% level. The 1-loop and the resummed tree level
decay agree at low scales where the resummation effects of the large
logarithms can be neglected, whereas the deviations are large for high
scales. In summary, in order to get correct predictions for the
flavour changing light stop decay for large scales of MFV, resummation
effects have to be included. To further improve on this decay, the
next step is the calculation of the one-loop corrections to the tree
level stop decay including the squark mixing matrix elements from RGE
evolution. This is deferred to a future publication. \s


\section{\label{sec:summary} Summary and Conclusions} 
In summary, we have calculated the flavour violating decay
$\tilde{t}_1 \to c \tilde{\chi}_1^0$ in the framework of Minimal
Flavour Violation including also finite terms, which do not depend on
the logarithm of the MFV scale $\mu_{\mbox{\scriptsize MFV}}$. The
one-loop decay has been compared to the approximate result derived
earlier by Hikasa and Kobayashi which neglects the subleading terms
compared to the large logarithm of $\mu_{\mbox{\scriptsize MFV}}$. It
has been found 
that it approximates the complete one-loop result within 10\% for large MFV
scales. The approximation becomes worse with decreasing scales. The
one-loop result, and also the approximate formula, however, do not
resum the large logarithms. The resummation is done by solving the
renormalization group equations. Since MFV is not RG-invariant, flavour
changing off-diagonal elements are induced in the squark mixing
matrices which lead to FCNC couplings at tree level. They can be
compared to the effective one-loop coupling in the MFV approach. The
resummation effects turn out to be important, so that the one-loop
result and the formula by Hikasa and Kobayashi only give an
approximate value of the phenomenologically important light stop decay
width into charm and neutralino. The next important step to improve
the prediction for the light stop decay width will be the calculation
of the one-loop corrections to the flavour-violating tree level decay.  

\section*{Appendix}
\section*{A Couplings}
To set up our notation for the couplings, we briefly repeat the
chargino and neutralino systems. The chargino mass matrix, in terms of
the wino mass parameter $M_2$, $\mu$ and $\tan\beta$, is given by
\cite{gunionhaber} 
\beq
{\cal M}_C = \left( \begin{array}{cc} M_2 & \sqrt{2} M_W s_\beta \\
\sqrt{2} M_W c_\beta & \mu \end{array} \right) \;,
\eeq
where $M_W$ denotes the charged $W$ boson mass and we use $s_\beta \equiv
\sin \beta$, $c_\beta \equiv \cos\beta$. It is diagonalized by two real
matrices $U$ and $V$,
\beq
U^* {\cal M}_C V^{-1} \quad \to \quad U = {\cal O}_- \quad
\mbox{and} \quad V = \left\{ \begin{array}{cl} {\cal O}_+  & \quad
    \mbox{if } \det {\cal M}_C > 0 \\ \sigma_3
    {\cal O}_+   & \quad \mbox{if } \det {\cal M}_C < 0 \end{array}
\right. \; ,
\eeq
with the Pauli matrix $\sigma_3$ rendering the chargino masses
positive. ${\cal O}_{\pm}$ are rotation matrices with the mixing
angles
\beq
\tan 2\theta_- = \frac{2\sqrt{2} M_W (M_2 c_\beta + \mu
  s_\beta)}{M_2^2-\mu^2-2M_W^2 c_\beta} \quad , \quad
\tan 2\theta_+ = \frac{2\sqrt{2} M_W (M_2 s_\beta + \mu
  c_\beta)}{M_2^2-\mu^2+2M_W^2 c_\beta} \; .
\eeq
The two chargino masses read
\beq
m^2_{\chi^\pm_{1,2}} = \frac{1}{2} \left\{ M_2^2 +\mu^2+2M_W^2 \mp
  [(M_2^2-\mu^2)^2+4M_W^2 (M_W^2 c_{2\beta}^2 + M_2^2 + \mu^2 + 2M_2
  \mu s_{2\beta})]^{\frac{1}{2}} \right\} \;.
\eeq 
The four-dimensional neutralino mass matrix in the
$(-i\tilde{B},-i\tilde{W}_3,\tilde{H}_1^0,\tilde{H}_2^0)$ basis has
the form
\beq
{M}_N= \left( \begin{array}{cccc} M_1 & 0 & -M_Z s_W c_\beta & 
M_Z s_W s_\beta \\ 0 & M_2 & M_Z c_W c_\beta & -M_Z c_W s_\beta \\
-M_Z s_W c_\beta & M_Z c_W c_\beta & 0 & -\mu \\ M_Z s_W s_\beta &
-M_Z c_W s_\beta & -\mu & 0  \end{array} \right) \;,
\eeq
with $c_W^2 = 1-s_W^2 = M_W^2/M_Z^2$. It can be diagonalized
analytically \cite{ElKheishen:1992yv} with a single matrix $Z$. \s

In the following, we list the couplings in the framework of MFV
\cite{gunionhaber,haberkane,Djouadi:1996pj,Djouadi:2001fa}, 
which are
needed for our calculation. All couplings are normalized to the weak
gauge coupling $g$ if not stated otherwise. Note that all charged
couplings involving quarks and/or squarks have to be multiplied with the CKM
matrix element ${\cal V}_{ud}$, which we have factored out from our
definition of the couplings.\s

$\bullet$ The couplings of charginos and neutralinos to the charged
gauge bosons $W^\pm$: 
\beq
G^{L,R}_{\tilde\chi_i^0\tilde\chi_j^+ W^+} = G^{L,R}_{ijW^+} \quad
\mbox{with} \quad \begin{array}{ccc} G^L_{ijW^+} &=& \frac{1}{\sqrt{2}} 
[-Z_{i4} V_{j2} + \sqrt{2} Z_{i2} V_{j1}] \\
G^R_{ijW^+} &=& \frac{1}{\sqrt{2}} [Z_{i3} U_{j2} + \sqrt{2} Z_{i2} U_{j1}]
\end{array}
\eeq

$\bullet$ The couplings of charginos and neutralinos to charged Higgs
and Goldstone bosons\footnote{We work in the
  Feynman gauge.},\\
\hspace*{0.8cm} $H^\pm$, $G^\pm$:
\beq
G^{L,R}_{\tilde{\chi}^0_i \tilde{\chi}_j^+ H^+} = G^{L,R}_{ijH^+} \quad
\mbox{with} \quad \begin{array}{ccc} G^L_{ijH^+} &=& c_\beta
  [Z_{i4} V_{j1} + \frac{1}{\sqrt{2}} (Z_{i2} + \tan\theta_W Z_{i1})
  V_{j2}] 
\\ 
G^R_{ijH^+} &=& s_\beta
  [Z_{i3} U_{j1} - \frac{1}{\sqrt{2}} (Z_{i2} + \tan\theta_W Z_{i1})
  U_{j2}] \end{array}
\eeq
\beq
G^{L,R}_{\tilde{\chi}^0_i \tilde{\chi}_j^+ G^+} = G^{L,R}_{ijG^+} \quad
\mbox{with} \quad \begin{array}{ccc} G^L_{ijG^+} &=& s_\beta
  [Z_{i4} V_{j1} + \frac{1}{\sqrt{2}} (Z_{i2} + \tan\theta_W Z_{i1})
  V_{j2}] 
\\ 
G^R_{ijG^+} &=& -c_\beta
  [Z_{i3} U_{j1} - \frac{1}{\sqrt{2}} (Z_{i2} + \tan\theta_W Z_{i1})
  U_{j2}] \end{array}
\eeq

$\bullet$ The couplings between neutralinos, quarks and squarks,
$\tilde{q}_{1,2}-q-\tilde{\chi}_j^0$:
\beq
\left\{ \begin{array}{c} a_{j1}^q \\ a_{j2}^q  \end{array} \right\}
&=& -\frac{m_q r_q}{\sqrt{2} M_W} \left\{ \begin{array}{c}  
    s_{\theta_q} \\ c_{\theta_q} \end{array} \right\} - e_{Lj}^q
\left\{ \begin{array}{c} c_{\theta_q} \\ -s_{\theta_q} \end{array}
\right\}
\nonumber \\
\left\{ \begin{array}{c} b_{j1}^q \\ b_{j2}^q  \end{array} \right\}
&=& -\frac{m_q r_q}{\sqrt{2} M_W} \left\{ \begin{array}{c}  
    c_{\theta_q} \\ -s_{\theta_q} \end{array} \right\} - e_{Rj}^q
\left\{ \begin{array}{c} s_{\theta_q} \\ c_{\theta_q} \end{array}
\right\} \; ,
\label{eq:coupneutqsq}
\eeq
with $r_u=Z_{j4}/\sin\beta$ and $r_d=Z_{j3}/\cos\beta$ for up- and
down-type quarks, and
\beq
e_{Lj}^q &=& \sqrt{2} \left[ Z_{j1} t_W (Q_q-I^3_q) +
  Z_{j2} I_q^3 \right] \nonumber \\
e_{Rj}^q &=& -\sqrt{2} Q_q t_W Z_{j1} \; ,
\label{eq:eler}
\eeq
where $t_W\equiv \tan\theta_W$.\s

$\bullet$ The couplings between charginos, quarks and squarks,
$\tilde{q}_{1,2}-q'-\tilde{\chi}_j^+$, for up- and down-type \\
\hspace*{0.8cm} (s)quarks read
\beq
\left\{ \begin{array}{c} a_{j1}^{\tilde{u} d} \\
    a_{j2}^{\tilde{u} d} \end{array} \right\} &=& V_{j1}
\left\{ \begin{array}{c} -c_{\theta_u} \\ s_{\theta_u} \end{array}
\right\} + \frac{m_u V_{j2} }{\sqrt{2} M_W s_\beta} 
\left\{ \begin{array}{c} s_{\theta_u} \\ c_{\theta_u} \end{array}
\right\} \nonumber \\
\left\{ \begin{array}{c} b_{j1}^{\tilde{u} d} \\
    b_{j2}^{\tilde{u} d} \end{array} \right\} &=& \frac{m_d
  U_{j2}}{\sqrt{2} M_W c_\beta}
\left\{ \begin{array}{c} c_{\theta_u}  \\ -s_{\theta_u} \end{array}   
\right\}
\label{eq:coup1}
\eeq
\beq
\left\{ \begin{array}{c} a_{j1}^{\tilde{d} u} \\
    a_{j2}^{\tilde{d} u} \end{array} \right\} &=& U_{j1}
\left\{ \begin{array}{c} -c_{\theta_d} \\ s_{\theta_d} \end{array}
\right\} + \frac{m_d U_{j2}}{\sqrt{2} M_W c_\beta} 
\left\{ \begin{array}{c} s_{\theta_d} \\ c_{\theta_d} \end{array}
\right\} \nonumber \\
\left\{ \begin{array}{c} b_{j1}^{\tilde{d} u} \\
    b_{j2}^{\tilde{d}u } \end{array} \right\} &=& \frac{m_u
  V_{j2}}{\sqrt{2} M_W s_\beta}
\left\{ \begin{array}{c} c_{\theta_d}  \\ -s_{\theta_d} \end{array}   
\right\}
\label{eq:coup2}
\eeq

$\bullet$ The couplings of the $W^\pm$ gauge bosons, the charged Higgs
and Goldstone bosons to quarks\\
\hspace*{0.8cm} are given by
\beq
v_W^q = -a_W^q = -\frac{1}{2\sqrt{2}} 
\eeq
\beq
v_{H^+}^q = \frac{m_d \tan\beta+ m_u \cot\beta}{2\sqrt{2} M_W}
\; , \; 
a_{H^+}^q = \frac{m_d \tan\beta- m_u \cot\beta}{2\sqrt{2} M_W}
\eeq
\beq
v_{G^+}^q = \frac{-m_d+ m_u}{2\sqrt{2} M_W}
\; , \; 
a_{G^+}^q = -\frac{m_d + m_u}{2\sqrt{2} M_W} \; .
\eeq 

$\bullet$ The couplings of the $W^\pm$ gauge bosons and charged Higgs
and Goldstone bosons to squarks \\ 
\hspace*{0.8cm} are given by
\beq
G_{W^+\tilde{u}_i\tilde{d}_j} = -\frac{1}{\sqrt{2}} C_{W^+ij} \qquad \mbox{with}
\nonumber \\[0.1cm]
\begin{array}{llllll} C_{W^+11} &=& c_{\theta_u} c_{\theta_d} & 
 C_{W^+12} &=& -c_{\theta_u} s_{\theta_d}  \\
 C_{W^+21} &=& -s_{\theta_u} c_{\theta_d} &
 C_{W^+22} &=& s_{\theta_u} s_{\theta_d} \end{array} 
\eeq
\beq
G_{H^+\tilde{u}_i\tilde{d}_j} &=& C_{H^+ij} \qquad \mbox{with}
\nonumber \\[0.1cm]
C_{H^+ 11} &=& c_{\theta_u} c_{\theta_d} s_{11} + s_{\theta_u}
s_{\theta_d} s_{22} + c_{\theta_u} s_{\theta_d} s_{12} + s_{\theta_u}
c_{\theta_d} s_{21} \nonumber \\
C_{H^+ 22} &=& s_{\theta_u} s_{\theta_d} s_{11} + c_{\theta_u}
c_{\theta_d} s_{22} - s_{\theta_u} c_{\theta_d} s_{12} - c_{\theta_u}
s_{\theta_d} s_{21}  \\
C_{H^+ 12} &=& -c_{\theta_u} s_{\theta_d} s_{11} + s_{\theta_u}
c_{\theta_d} s_{22} + c_{\theta_u} c_{\theta_d} s_{12} - s_{\theta_u}
s_{\theta_d} s_{21} \nonumber \\
C_{H^+ 21} &=& -s_{\theta_u} c_{\theta_d} s_{11} + c_{\theta_u}
s_{\theta_d} s_{22} - s_{\theta_u} s_{\theta_d} s_{12} + c_{\theta_u}
c_{\theta_d} s_{21} \;, \nonumber
\eeq
where
\beq
s_{11} &=& -\frac{M_W}{\sqrt{2}} \left( \sin 2\beta - \frac{m_d^2
    \tan\beta + m_u^2 \cot\beta}{M_W^2} \right) \nonumber \\
s_{22} &=& \frac{m_u m_d}{\sqrt{2}M_W} (\tan\beta+\cot\beta) \\
s_{12} &=& \frac{m_d}{\sqrt{2} M_W} (\mu+A_d \tan\beta) \nonumber \\
s_{21} &=& \frac{m_u}{\sqrt{2} M_W} (\mu+A_u \cot\beta) \nonumber
\; .
\eeq
And for the Goldstone couplings we have
\beq
G_{G^+\tilde{u}_i\tilde{d}_j} &=& C_{G^+ij} \qquad \mbox{with}
\nonumber \\[0.1cm]
C_{G^+11} &=& c_{\theta_u} c_{\theta_d} a_{LL} + c_{\theta_u}
s_{\theta_d} a_{LR} + s_{\theta_u} c_{\theta_d} a_{RL} \nonumber \\
C_{G^+22} &=& s_{\theta_u} s_{\theta_d} a_{LL} - s_{\theta_u}
c_{\theta_d} a_{LR} - c_{\theta_u} s_{\theta_d} a_{RL} \\
C_{G^+12} &=& -c_{\theta_u} s_{\theta_d} a_{LL} + c_{\theta_u}
c_{\theta_d} a_{LR} - s_{\theta_u} s_{\theta_d} a_{RL} \nonumber \\
C_{G^+21} &=& -s_{\theta_u} c_{\theta_d} a_{LL} - s_{\theta_u}
s_{\theta_d} a_{LR} + c_{\theta_u} c_{\theta_d} a_{RL} \; , \nonumber
\eeq
and
\beq
a_{LL} &=& \frac{M_W}{\sqrt{2}} \left( \cos 2\beta +
  \frac{m_u^2-m_d^2}{M_W^2} \right) \nonumber \\
a_{LR} &=& \frac{m_d}{\sqrt{2} M_W} (\mu \tan \beta - A_d) \\
a_{RL} &=& -\frac{m_u}{\sqrt{2} M_W} (\mu \cot\beta - A_u)  \nonumber
\; .
\eeq
\s

$\bullet$ For our calculation we also need the 4-squark coupling between stop,
scharm and two {\it identical} down-type squarks. With the generation index
$k=1,2,3$ and the index $l=1,2$ denoting the two down-type squark
eigenstates, it reads in terms of $g^2$
\beq
G_{\tilde{t}_i \tilde{c}_j \tilde{d}_{kl} \tilde{d}_{kl}} = C_{ij
  \tilde{d}_{kl} \tilde{d}_{kl}}  \quad
\mbox{with} \quad
C_{ij \tilde{d}_{kl} \tilde{d}_{kl}} = -\frac{m_t m_c}{2 M_W^2
  s_\beta^2} {\cal P}_{ijll} - 
\frac{m_{d_k}^2}{2 M_W^2 c_\beta^2} {\cal P}'_{ijll} - \frac{1}{2} {\cal
  P}^{''}_{ijll}
\eeq
and
\beq
\renewcommand{\arraystretch}{1.15}
\begin{array}{|c|c|c|c|}  \hline
ijll & {\cal P}_{ijll} & {\cal P}'_{ijll} & {\cal P}^{''}_{ijll} \\ \hline
1111 & s_{\theta_t} s_{\theta_c} c_{\theta_{d_k}}^2 & c_{\theta_t}
c_{\theta_c} s_{\theta_{d_k}}^2 &  c_{\theta_c} c_{\theta_t}
c_{\theta_{d_k}}^2 
\\ \hline
2211 & c_{\theta_t} c_{\theta_c} c_{\theta_{d_k}}^2 & s_{\theta_c}
s_{\theta_t} s_{\theta_{d_k}}^2 & s_{\theta_c} s_{\theta_t} c_{\theta_{d_k}}^2
\\ \hline
1211 & s_{\theta_t} c_{\theta_c} c_{\theta_{d_k}}^2 & -c_{\theta_t}
s_{\theta_c} s_{\theta_{d_k}}^2 & -c_{\theta_t} s_{\theta_c}
c_{\theta_{d_k}}^2 \\ \hline
2111 & c_{\theta_t} s_{\theta_c} c_{\theta_{d_k}}^2 & - c_{\theta_c}
s_{\theta_t} s_{\theta_{d_k}}^2 & -c_{\theta_c} s_{\theta_t}
c_{\theta_{d_k}}^2 \\ \hline
\end{array}
\eeq
The couplings with two $\tilde{d}_{k2}$ squarks are obtained from those
with the $\tilde{d}_{k1}$ squarks by interchanging $c_{\theta_{d_k}}
\leftrightarrow s_{\theta_{d_k}}$. \s

$\bullet$ Finally, the couplings between stop, scharm and two charged
Higgs bosons or two charged Goldstone bosons in terms of $g^2$ are
given by
\beq
G_{\tilde{t}_i \tilde{c}_j H^+ H^+} = 
-\frac{m_b^2\tan\beta^2}{2M_W^2} Q_{ijH^+ H^+}
\eeq
and
\beq
G_{\tilde{t}_i \tilde{c}_j G^+G^+} = -\frac{m_b^2}{2M_W^2} Q_{ijH^+ H^+}
\eeq
where
\beq
\begin{array}{llllll} Q_{11H^+ H^+} &=& c_{\theta_t} c_{\theta_c} & 
 Q_{22H^+ H^+} &=& s_{\theta_t} s_{\theta_c}  \\
 Q_{12H^+ H^+} &=& -c_{\theta_t} s_{\theta_c} &
 Q_{21H^+ H^+} &=& -s_{\theta_t} c_{\theta_c} \end{array} \; .
\eeq

\section*{\label{subsec:squark} B1 Squark self-energy contributions} 
In this Appendix we give the result for the self-energy
$\tilde{\Sigma}_{\tilde{c}_L \tilde{t}_1}$ in terms of the couplings
defined in Appendix A.  The self-energy receives contributions from
the various diagrams in Fig.\ref{fig:alldiags}, {\it
  i.e.}
\beq
\tilde{\Sigma}_{\tilde{c}_L \tilde{t}_1} = \Sigma^{\tilde{\chi}^+
  d}_{\tilde{c}_L \tilde{t}_1} 
+ \Sigma^{H^+ \tilde{d}}_{\tilde{c}_L \tilde{t}_1}
+ \Sigma^{G^+ \tilde{d}}_{\tilde{c}_L \tilde{t}_1}
+ \Sigma^{W^+ \tilde{d}}_{\tilde{c}_L \tilde{t}_1}
+ \Sigma^{H^+ G^+}_{\tilde{c}_L \tilde{t}_1}
+ \Sigma^{4\tilde{q}}_{\tilde{c}_L \tilde{t}_1} \;.
\eeq
Note that the self-energy for $\tilde{t}_1$ and $\tilde{c}_R$ external
legs is zero for vanishing $c$ quark mass, 
\beq
\tilde{\Sigma}_{\tilde{c}_R \tilde{t}_1} |_{m_c=0}  = 0 \;.
\eeq
We have for
\beq
\Sigma^{\tilde{\chi}^+ d}_{\tilde{c}_L \tilde{t}_1}
(m_{\tilde{t}_1}^2) &=& g^2 \sum_{j=1,2} \sum_{k=1,..,3} 
\, {\cal V}_{cd_k} {\cal V}_{td_k}^* \,  (-2) \,
\Big\{ \frac{1}{2} ( \acdj1 \atdj1 + \bcdj1 \btdj1 )
  [A_0 (\mchij^2) + A_0 (\mdk^2) \nonumber \\
&+& (\mchij^2 -\mst1^2 + \mdk^2) B_0
  (\mst1^2,\mchij,\mdk)] \nonumber \\ 
&+& (\acdj1 \btdj1 + \atdj1 \bcdj1) \mchij \mdk B_0
  (\mst1^2,\mchij,\mdk) \Big\}  \;,
\eeq
with $\acdj1$ etc. given in Eq.~(\ref{eq:coup1}). 
The sum is taken over the chargino eigenstates and the three quark
generations. The scalar
one-loop one- and two-point integrals $A_0(m)$ and $B_0(p^2,m_1,m_2)$
are defined as \cite{loopintegs} 
\beq
A_0 (m) &=& -i\, \bar{\mu}^{4-n} \int \frac{d^n k}{(2\pi)^n} \frac{1}{k^2-m^2}
\\
B_0 (p^2,m_1,m_2) &=& -i\, \bar{\mu}^{4-n} \int \frac{d^n k}{(2\pi)^n}
\frac{1}{(k^2-m_1^2)[(k+p)^2-m_2^2]} \; .
\eeq
Note the suppression by the CKM matrix elements ${\cal V}_{cd_k}$,
${\cal V}_{td_k}^*$.  We find for $\Sigma^{H^+ \tilde{d}}_{\tilde{c}_L \tilde{t}_1}$: 
\beq
\Sigma^{H^+ \tilde{d}}_{\tilde{c}_L \tilde{t}_1}
(m_{\tilde{t}_1}^2) &=& g^2 
\sum_{k=1,..,3} \, \sum_{i=1,2} 
\, {\cal V}_{cd_k} {\cal V}_{td_k}^* \, 
G_{H^+\tilde{t}_1 \tilde{d}_{ki}} G_{H^+\tilde{c}_L \tilde{d}_{ki}}
B_0 (\mst1^2,m_{H^+},m_{\tilde{d}_{ki}}) \; .
\eeq
The sum is to be taken over the three squark generations $k$ and the
two squark mass eigenstates $i$. The Goldstone contribution reads
\beq
\Sigma^{G^+ \tilde{d}}_{\tilde{c}_L \tilde{t}_1}
(m_{\tilde{t}_1}^2) &=& g^2 
\sum_{k=1,..,3} \, \sum_{i=1,2} 
\, {\cal V}_{cd_k} {\cal V}_{td_k}^* \, 
G_{G^+\tilde{t}_1 \tilde{d}_{ki}} G_{G^+\tilde{c}_L \tilde{d}_{ki}}
B_0 (\mst1^2,M_W,m_{\tilde{d}_{ki}}) \;, 
\eeq
and the self-energy involving the $W$ boson
\beq
\Sigma^{W^+ \tilde{d}}_{\tilde{c}_L \tilde{t}_1}
(m_{\tilde{t}_1}^2) &=& g^2 \, 
\sum_{k=1,..,3} \, \sum_{i=1,2} 
\, {\cal V}_{cd_k} {\cal V}_{td_k}^* \, G_{W^+ \tilde{t}_1
  \tilde{d}_{ki}} G_{W^+ \tilde{c}_L \tilde{d}_{ki}} \Big\{-2 A_0
(M_W^2) + A_0 (m_{\tilde{d}_{ki}}^2) \nonumber \\
&-& (2 \mst1^2 + 2 \mdki^2 - M_W^2)
B_0 (\mst1^2,M_W,\mdki) \Big\} \;.
\eeq
Finally, we have the tadpole contributions from the charged Higgs and
Goldstone boson loop, 
\beq
\Sigma^{H^+ G^+}_{\tilde{c}_L \tilde{t}_1}
(m_{\tilde{t}_1}^2) &=& g^2 (-1) \, {\cal V}_{cb} {\cal V}_{tb}^* 
\left[G_{\tilde{t}_1 \tilde{c}_L H^+H^+} A_0 (m_{H^+}^2)+ 
G_{\tilde{t}_1 \tilde{c}_L G^+G^+} A_0(M_W^2) \right] \; ,
\eeq
and the one from the 4-squark vertex
\beq
\Sigma^{4\tilde{q}}_{\tilde{c}_L \tilde{t}_1}
(m_{\tilde{t}_1}^2) &=& g^2 \, 
\sum_{k=1,..,3} \, \sum_{i=1,2} \, {\cal V}_{cd_k} {\cal V}_{td_k}^* 
G_{\tilde{t}_1 \tilde{c}_L \tilde{d}_{ki} \tilde{d}_{ki}} A_0
(\mdki^2) \; .
\eeq

\section*{\label{subsec:quark} B2 Quark self-energy} 
According to the structure of the quark self-energy given in
Eq.~(\ref{eq:strucself}) we find for the left-chiral contribution
\beq
\Sigma^L_{tc} (p^2) &=& g^2 \Big\{ {\cal V}_{cb} {\cal V}_{tb}^*
\frac{m_b^2}{2 M_W^2} \left[ -t^2_\beta B_1 (p^2,m_b,m_{H^+}) -
  B_1   (p^2,m_b,M_W)\right] \nonumber \\
&& + \sum_{j=1,2} \sum_{k=1,..,3}
\sum_{i=1,2} {\cal V}_{cd_k} {\cal V}_{td_k}^*  (-a^{\tilde{d}_k c}_{ji}
a^{\tilde{d}_k t}_{ji} ) 
B_1(p^2,\mchij,m_{\tilde{d}_{ki}}) \nonumber \\ 
&& +\sum_{k=1,..,3} \frac{1}{2} {\cal V}_{cd_k} {\cal V}_{td_k}^* [-1-2B_1
(p^2,m_{d_k},M_W)] \Big\} \;,
\eeq
with $a^{\tilde{d}_k c}_{ji}, a^{\tilde{d}_k t}_{ji}$ given in Eq.~(\ref{eq:coup2}). 
The sums are taken over all possible chargino states ($j=1,2$), the
three quark and squark generations ($k=1,2,3$) and the two squark mass
eigenstates ($i=1,2$). Furthermore, $B_1$ in terms of the scalar one-
and two-point functions is given by
\beq
B_1 (p^2,m_0,m_1) = \frac{1}{2p^2} \left[ A_0 (m_0) - A_0 (m_1) -
  (p^2-m_1^2+m_0^2) B_0 (p^2,m_0,m_1) \right] \;.
\eeq
The right-chiral contribution reads
\beq
\Sigma^R_{tc} (p^2) &=& g^2 \Big\{ \sum_{k=1,..,3} {\cal V}_{cd_k} 
{\cal V}_{td_k}^* \frac{m_c m_t}{2 M_W^2} \left[ -\cot^2_\beta B_1
  (p^2,m_{d_k},m_{H^+}) -  B_1   (p^2,m_{d_k},M_W)\right] \nonumber \\
&& + \sum_{j=1,2} \sum_{k=1,2,3}
\sum_{i=1,2} {\cal V}_{cd_k} {\cal V}_{td_k}^*  (-b^{\tilde{d}_k c}_{ji}
b^{\tilde{d}_k t}_{ji} ) B_1(p^2,\mchij,m_{\tilde{d}_{ki}}) \Big\} \;,
\eeq
with $b^{\tilde{d}_k c}_{ji}, b^{\tilde{d}_k t}_{ji}$ defined in Eq.~(\ref{eq:coup2}).
It vanishes for $m_c=0$. The left-chiral scalar contribution can be
cast into the form
\beq
\Sigma^{Ls}_{tc} (p^2) &=& g^2 \Big\{ {\cal V}_{cb} {\cal V}_{tb}^*
\frac{m_b^2 m_c}{2M_W^2} [B_0(p^2,m_{H^+},m_b)-B_0(p^2,M_W,m_b)] 
\nonumber \\
&& + \sum_{j=1,2} \sum_{k=1,..,3} \sum_{i=1,2} {\cal V}_{cd_k} {\cal
  V}_{td_k}^* \mchij a^{\tilde{d}_kt}_{ji} b^{\tilde{d}_kc}_{ji}
B_0(p^2,m_{\tilde{d}_{ki}},\mchij) \Big\} \;,
\eeq
which also vanishes for zero charm quark mass. For the right-chiral
scalar contribution we find 
\beq
\Sigma^{Rs}_{tc} (p^2) &=& g^2 \Big\{ {\cal V}_{cb} {\cal V}_{tb}^*
\frac{m_b^2 m_t}{2M_W^2} [B_0(p^2,m_{H^+},m_b)-B_0(p^2,M_W,m_b)] 
\nonumber \\
&& + \sum_{j=1,2} \sum_{k=1,..,3} \sum_{i=1,2} {\cal V}_{cd_k} {\cal
  V}_{td_k}^* \mchij b^{\tilde{d}_kt}_{ji} a^{\tilde{d}_kc}_{ji}
B_0(p^2,m_{\tilde{d}_{ki}},\mchij) \Big\} \;.
\eeq
Note that in case of real two-point functions we have
$\Sigma^{Rs}_{tc}= \Sigma^{Ls*}_{ct}$.

\section*{B3 Vertex correction}
The vertex contributions to the left-chiral form factor $F^v_L$
vanish for $m_c = 0$. For the right-chiral form factor $F^v_R$ they
are given by the various right-chiral contributions from the vertex correction
graphs depicted in Fig.\ref{fig:alldiags} 
\beq
g F_{R}^v (\mst1^2) &=&  i\, [\Gamma_{\tilde{\chi}^+ \tilde{d} d} +
\Gamma_{\tilde{\chi}^+ H^+ d} + \Gamma_{\tilde{\chi}^+ G^+ d} +
\Gamma_{\tilde{\chi}^+ W^+ d}  + \Gamma_{\tilde{d} H^+ d}   +
\Gamma_{\tilde{d} G^+ d} + \Gamma_{\tilde{\chi}^+ \tilde{d} H^+} +
\Gamma_{\tilde{\chi}^+ \tilde{d} G^+}  \nonumber \\
&& + \Gamma_{\tilde{d} d W^+} + \Gamma_{\tilde{d} \tilde{\chi}^+
  W^+}](\mst1^2)  \; .
\eeq
We have for $\Gamma_{\tilde{\chi}^+ \tilde{d} d}$,
\beq
\Gamma_{\tilde{\chi}^+ \tilde{d} d} (\mst1^2) &=& 
- g^3 \sum_{j=1,2} \sum_{k=1...3}
\sum_{i=1,2}  {\cal V}_{cd_k} {\cal V}_{td_k}^* \Big\{ [c_{1ijk} \mn1
+c_{2ijk} \mst1^2 ] B_0 (\mst1^2,m_{d_k},\mchij) \nonumber \\
&-& [c_{1ijk} \mn1  +c_{2ijk} \mn1^2] 
B_0(\mn1^2,m_{\tilde{d}_{ki}},m_{d_k})   
 + [ c_{1ijk} \mn1  ( \mst1^2-\mn1^2 + m_{\tilde{d}_{ki}}^2 -
 \mchij^2) \nonumber \\
& & -c_{2ijk}  (\mchij^2 \mn1^2 -\mst1^2 m_{\tilde{d}_{ki}}^2) +(-
c_{3ijk} \mn1 \mdk+c_{4ijk} \mchij \mdk) (\mst1^2-\mn1^2)  ] \nonumber \\
&& C_0 (\mst1^2,\mn1^2,0,\mchij,\mdk,m_{\tilde{d}_{ki}})
\Big\}/(\mst1^2-\mn1^2) \;,
\eeq
where the sum over all
possible chargino eigenstates $\tilde{\chi}_j^+$ ($j=1,2$), all three
generations of down type quarks and squarks ($k=1,2,3$) as well as the
two squark mass eigenstates ($i=1,2$) has to be taken. We have
introduced the abbreviations 
\beq
\begin{array}{lcllcl}
c_{1ijk} &=& a^{\tilde{d}_k c}_{ji} a^{\tilde{t} d_k}_{j1} a^{d_k}_{1i}
m_{\tilde{\chi}^+_j}  + a^{\tilde{d}_k c}_{ji} b^{\tilde{t} d_k}_{j1}
a^{d_k}_{1i} \mdk & \qquad
c_{2ijk} &=& a^{\tilde{d}_k c}_{ji} b^{\tilde{t} d_k}_{j1}
b^{d_k}_{1i} \\ 
c_{3ijk} &=& a^{\tilde{d}_k c}_{ji} b^{\tilde{t} d_k}_{j1}
a^{d_k}_{1i} & \qquad
c_{4ijk} &=& a^{\tilde{d}_k c}_{ji} a^{\tilde{t} d_k}_{j1}
b^{d_k}_{1i} \;,
\end{array} 
\eeq
with the various couplings defined in 
Eqs.~(\ref{eq:coupneutqsq},\ref{eq:coup1},\ref{eq:coup2}).
 The scalar one-loop 3-point function is given by
\beq
C_0 (p_1^2,p_2^2,(p_1+p_2)^2,m_1,m_2,m_3) &=& \nonumber
\\ -i\, \bar{\mu}^{4-n} \int && \hspace*{-0.9cm} \frac{d^n
  k}{(2\pi)^n}
\frac{1}{(k^2-m_1^2)[(k+p_1)^2-m_2^2][(k+p_1+p_2)^2-m_3^2]} \; . 
\eeq
We find
\beq
\Gamma_{\tilde{\chi}^+ H^+ d} (\mst1^2) &=& 
g^3 \sum_{j=1,2} {\cal V}_{cb} {\cal V}_{tb}^*
\frac{m_b \tan\beta}{\sqrt{2} M_W} \Big\{ [c_{1j} \mn1 + c_{2j}
\mst1^2] B_0 (\mst1^2,\mchij,m_b) \nonumber\\
&-& [c_{1j} \mn1 + c_{2j} \mn1^2]
B_0 (\mn1^2,m_{H^+},\mchij) + [c_{1j} \mn1
(\mst1^2-\mn1^2-m_b^2+m_{H^+}^2) \nonumber\\
&&+ c_{2j} (\mst1^2
m_{H^+}^2-\mn1^2 m_b^2) + \mchij (c_{3j} m_b - c_{4j} \mn1) (\mst1^2-
\mn1^2)]
\nonumber \\
&& C_0 (\mst1^2,\mn1^2,0,m_b,\mchij,m_{H^+}) 
\Big\} /(\mst1^2-\mn1^2) \;,
\eeq
with
\beq
\begin{array}{lcllcl}
c_{1j} &=& a^{\tilde{t} b}_{j1} G^L_{1jH^+} m_b + b^{\tilde{t}
  b}_{j1} G^L_{1j4} m_{\tilde{\chi}_j^+} & \qquad
c_{2j} &=& b^{\tilde{t} b}_{j1} G^R_{1jH^+} \nonumber \\
c_{3j} &=& a^{\tilde{t} b}_{j1} G^R_{1jH^+} & \qquad
c_{4j} &=& b^{\tilde{t} b}_{j1} G^L_{1jH^+} \;.
\end{array}
\eeq
Note that we set the down and strange quark mass to zero,
$m_d=m_s=0$. We have 
\beq
\Gamma_{\tilde{\chi}^+ G^+ d} (\mst1^2) &=& 
g^3 \sum_{j=1,2} {\cal V}_{cb} {\cal V}_{tb}^*
\frac{-m_b}{\sqrt{2} M_W} \Big\{ [c_{5j} \mn1 + c_{6j}
\mst1^2] B_0 (\mst1^2,\mchij,m_b) \nonumber\\
&-& [c_{5j} \mn1 + c_{6j} \mn1^2]
B_0 (\mn1^2,M_{W},\mchij) + [c_{5j} \mn1
(\mst1^2-\mn1^2-m_b^2+M_{W}^2) \nonumber\\
&&+ c_{6j} (\mst1^2
M_{W}^2-\mn1^2 m_b^2) + \mchij (c_{7j} m_b - c_{8j} \mn1)
(\mst1^2-\mn1^2)]
\nonumber \\
&& C_0 (\mst1^2,\mn1^2,0,m_b,\mchij,M_W) 
\Big\} /(\mst1^2-\mn1^2) \;,
\eeq
where
\beq
\begin{array}{lcllcl}
c_{5j} &=& a^{\tilde{t} b}_{j1} G^L_{1jG^+} m_b + b^{\tilde{t}
  b}_{j1} G^L_{1jG^+} m_{\tilde{\chi}_j^+} & \qquad
c_{6j} &=& b^{\tilde{t} b}_{j1} G^R_{1jG^+} \nonumber \\
c_{7j} &=& a^{\tilde{t} b}_{j1} G^R_{1jG^+} & \qquad
c_{8j} &=& b^{\tilde{t} b}_{j1} G^L_{1jG^+} \;.
\end{array}
\eeq
And
\beq
\Gamma_{\tilde{\chi}^+ W^+ d} (\mst1^2) &=& 
g^3 \sum_{j=1,2} \sum_{k=1...3}  {\cal V}_{cd_k} {\cal V}_{td_k}^* 
\frac{1}{\sqrt{2}} \Big\{ -2 c_{1jk} \mn1 B_0 (\mst1^2,\mchij,m_{d_k}) 
\nonumber \\
&+& 2 [c_{1jk} \mn1 + c_{2jk} (\mst1^2-\mn1^2)] B_0 (\mn1^2,M_W,\mchij) 
+ 2 c_{2jk} (\mst1^2-\mn1^2) \nonumber\\
&& B_0 (0,M_W,m_{d_k}) +
[2 c_{2jk} (m_{d_k}^2+\mchij^2-\mst1^2)(\mst1^2-\mn1^2) +2 \mn1 c_{1jk}
\nonumber\\
&& (\mn1^2-\mst1^2+m_{d_k}^2-M_W^2) +2 \mchij (c_{3jk} \mn1 + 2 c_{4jk}
m_{d_k}) (\mst1^2-\mn1^2)] \nonumber \\
&& C_0 (\mst1^2,\mn1^2,0,m_{d_k},\mchij,M_W) 
-2 c_{2jk} (\mst1^2-\mn1^2)
\Big\} /(\mst1^2-\mn1^2) \;,
\eeq
with
\beq
\begin{array}{lcllcl}
c_{1jk} &=& b^{\tilde{t} d_k}_{j1} G^R_{1jW^+} m_{d_k} + a^{\tilde{t}
  d_k}_{j1} G^R_{1jW^+} \mchij & \qquad
c_{2jk} &=& a^{\tilde{t} d_k}_{j1} G^L_{1jW^+} \\
c_{3jk} &=& a^{\tilde{t} d_k}_{j1} G^R_{1jW^+} & \qquad
c_{4jk} &=& b^{\tilde{t} d_k}_{j1} G^L_{1jW^+}  \; .
\end{array}
\eeq
Next
\beq
\Gamma_{\tilde{d} H^+ d} (\mst1^2) &=& 
g^3 \sum_{i=1,2}  {\cal V}_{cb} {\cal V}_{tb}^* \frac{m_b
  \tan\beta \, G_{H^+\tilde{t}_1\tilde{b}_i}}{\sqrt{2} M_W}  \Big\{
b^{b}_{1i} \mn1 
[B_0(\mst1^2,m_{\tilde{b}_i},m_H^+) -
B_0(\mn1^2,m_b,m_{\tilde{b}_i})] \nonumber \\
&-& [a^{b}_{1i} m_b (\mst1^2-\mn1^2)
- b^{b}_{1i} \mn1 (m_b^2-m_{H^+}^2)] C_0
(\mst1^2,\mn1^2,0,m_{H^+},m_{\tilde{b}_i},m_b) \Big\} \nonumber \\
&& /(\mst1^2-\mn1^2)
\eeq
and
\beq
\Gamma_{\tilde{d} G^+ d} (\mst1^2) &=& 
g^3 \sum_{i=1,2}  {\cal V}_{cb} {\cal V}_{tb}^* \frac{-m_b
  \, G_{G^+\tilde{t}_1\tilde{b}_i}}{\sqrt{2} M_W}  \Big\{ b^{b}_{1i} \mn1
[B_0(\mst1^2,m_{\tilde{b}_i},M_W) -
B_0(\mn1^2,m_b,m_{\tilde{b}_i})] \nonumber \\
&-& [a^{b}_{1i} m_b (\mst1^2-\mn1^2)
- b^{b}_{1i} \mn1 (m_b^2-M_W^2)]
C_0(\mst1^2,\mn1^2,0,M_W,m_{\tilde{b}_i},m_b) \Big\}  \nonumber \\
&& /(\mst1^2-\mn1^2) \;.
\eeq
Furthermore,
\beq
\Gamma_{\tilde{\chi}^+ \tilde{d} H^+} (\mst1^2) &=& 
g^3 \sum_{j=1,2} \sum_{k=1...3} \sum_{i=1,2} G_{H^+\tilde{t}_1
  \tilde{d}_{ki}}  {\cal V}_{cd_k} {\cal V}_{td_k}^* \Big\{ c_{5ijk}
\mn1 [-B_0(\mst1^2,m_{H^+},m_{\tilde{d}_{ki}}) \nonumber \\
&+& B_0(\mn1^2,\mchij,m_{H^+})] 
+ [c_{6ijk} \mchij (\mst1^2-\mn1^2) + c_{5ijk} \mn1
(m_{\tilde{d}_{ki}}^2-\mchij^2) ] \nonumber \\
&& C_0(\mst1^2,\mn1^2,0,m_{\tilde{d}_{ki}},m_{H^+},\mchij) \Big\}
/(\mst1^2-\mn1^2)  
\eeq
with
\beq
c_{5ijk} = G^R_{1jH^+} a^{\tilde{d}_k c}_{ji} \qquad \qquad 
c_{6ijk} = G^L_{1jH^+} a^{\tilde{d}_k c}_{ji} \;, 
\eeq
and
\beq
\Gamma_{\tilde{\chi}^+ \tilde{d} G^+} (\mst1^2) &=& 
g^3 \sum_{j=1,2} \sum_{k=1...3} \sum_{i=1,2} G_{G^+\tilde{t}_1
  \tilde{d}_{ki}}  {\cal V}_{cd_k} {\cal V}_{td_k}^* \Big\{ c_{7ijk}
\mn1 [-B_0(\mst1^2,M_W,m_{\tilde{d}_{ki}}) \nonumber \\
&+& B_0(\mn1^2,\mchij,M_W)] 
+ [c_{8ijk} \mchij (\mst1^2-\mn1^2) + c_{7ijk} \mn1
(m_{\tilde{d}_{ki}}^2-\mchij^2) ] \nonumber \\
&& C_0(\mst1^2,\mn1^2,0,m_{\tilde{d}_{ki}},M_W,\mchij) \Big\}
/(\mst1^2-\mn1^2)  \;,
\eeq
with
\beq
c_{7ijk} = G^R_{1jG^+} a^{\tilde{d}_k c}_{ji} \qquad \qquad 
c_{8ijk} = G^L_{1jG^+} a^{\tilde{d}_k c}_{ji} \;. 
\eeq
We have
\beq
\Gamma_{\tilde{d} d W^+} (\mst1^2) &=& 
g^3 \sum_{j=1,2} \sum_{k=1...3} \sum_{i=1,2} \frac{-G_{W^+\tilde{t}_1
  \tilde{d}_{ki}}}{\sqrt{2}}  {\cal V}_{cd_k} {\cal V}_{td_k}^* \Big\{
[-a^{d_k}_{1i} (\mst1^2+\mn1^2) - b^{d_k}_{1i} \mn1 m_{d_k}]
\nonumber\\ &&
B_0(\mst1^2,m_{\tilde{d}_{ki}},M_W) + [2 a^{d_k}_{1i} \mn1^2 +
b^{d_k}_{1i} \mn1 m_{d_k} ] B_0(\mn1^2,m_{d_k},m_{\tilde{d}_{ki}})
\nonumber\\
&+&  2 a^{d_k}_{1i} (\mst1^2-\mn1^2) B_0(0,m_{d_k},M_W)
+ [ 2 a^{d_k}_{1i} \big( \mn1^2 (\mn1^2-\mst1^2 - m_{\tilde{d}_{ki}}^2
-\frac{1}{2} m_{d_k}^2 + M_W^2) \nonumber \\
&& + \mst1^2 (m_{\tilde{d}_{ki}}^2 -
\frac{1}{2} m_{d_k}^2 ) \big)- b^{d_k}_{1i} \mn1 m_{d_k}
(2\mst1^2+m_{d_k}^2 - M_W^2-2\mn1^2)] \nonumber\\
&&C_0(\mst1^2,\mn1^2,0,M_W,m_{\tilde{d}_{ki}},m_{d_k})
\Big\}/(\mst1^2-\mn1^2)  \;.
\eeq
And finally
\beq
\Gamma_{\tilde{d} \tilde{\chi}^+ W^+} &=& g^3 \sum_{j=1,2}
\sum_{k=1...3} \sum_{i=1,2} G_{W^+\tilde{t}_1
    \tilde{d}_{ki}}  {\cal V}_{cd_k} {\cal V}_{td_k}^*
\Big\{ [c_{9ijk} \mst1^2 + c_{10ijk} \mn1 \mchij] 
B_0(\mst1^2,M_W,m_{\tilde{d}_{ki}}) \nonumber \\
&+& [c_{9ijk} (\mn1^2-2\mst1^2)-c_{10ijk} \mn1\mchij]
B_0(\mn1^2,\mchij,M_W) + [c_{9ijk}(\mst1^2 \mchij^2-2\mst1^2
m_{\tilde{d}_{ki}}^2 \nonumber \\
&& + \mn1^2 m_{\tilde{d}_{ki}}^2) + c_{10ijk}
\mn1\mchij (-\mst1^2+\mchij^2-m_{\tilde{d}_{ki}}^2+\mn1^2)]
\nonumber\\
&& C_0(\mst1^2,\mn1^2,0,m_{\tilde{d}_{ki}},M_W,\mchij)
\Big\}/(\mst1^2-\mn1^2) \;, 
\eeq
with
\beq
c_{9ijk} = G^R_{1jW^+} a^{\tilde{d}_k c}_{ji} \qquad \qquad
c_{10ijk} = G^L_{1jW^+} a^{\tilde{d}_k c}_{ji}  \;.
\eeq

\section*{C FCNC counterterm}
We start from  the $\tilde{u}-u-$neutralino part of
the Lagrangian in the interaction basis, expressed in terms of the
bare squark and quark fields, $\tilde{u}_i^{(0)}$ 
and $u_i^{(0)}$ and the bare quark mass matrix ${\bf m}^{(0)}_{ij}$, where
$i,j=1,2,3$ denote the generation indices and $l=1,..,4$ the
neutralino mass eigenstates,   
\beq
{\cal L}_{\bar{u}\tilde{u}\tilde{\chi}^0} &=& -\bar{u}^{(0)}_i \,
g\, e_{Ll}^{u_i} \, \tilde{u}_{iL}^{(0)} \, {\cal P}_R \, \tilde{\chi}^0_l +
\bar{u}^{(0)}_i \left( -\frac{g Z_{l4} {\bf m}^{(0)}_{ij}}{\sqrt{2} M_W
    \sin\beta}\right) \tilde{u}^{(0)}_{jR} \, {\cal P}_R \,
\tilde{\chi}^0_l \nonumber \\
&& -\bar{u}^{(0)}_i \,
g\, e_{Rl}^{u_i} \, \tilde{u}_{iR}^{(0)} \, {\cal P}_L \, \tilde{\chi}^0_l +
\bar{u}^{(0)}_i \left( -\frac{g Z_{l4} {\bf m}^{(0)}_{ij}}{\sqrt{2} M_W
    \sin\beta}\right) \tilde{u}^{(0)}_{jL} \, {\cal P}_L \,
\tilde{\chi}^0_l + h.c. \; .
\eeq
The couplings $e_{L,R \, l}^{u_i}$ have been defined in
Eq.~(\ref{eq:eler}). Let us look at the right-chiral part of the
coupling. Rotation to the mass eigenstates yields
\beq
{\cal L}_{\bar{u}\tilde{u}\tilde{\chi}^0}^R &=& -\bar{u}^{m(0)}_k
U^{u_L (0)}_{ki} \, g e_{Ll}^{u_i} \, \tilde{W}^{(0)\dagger}_{is} \tilde{u}^{m(0)}_{s}
\, {\cal P}_R \, \tilde{\chi}^0_l +
\bar{u}^{m(0)}_k U^{u_L (0)}_{ki} 
\frac{-g Z_{l4} {\bf m}^{(0)}_{ij}}{\sqrt{2} M_W
    \sin\beta} \tilde{W}^{(0) \dagger}_{j+3 \, s} \tilde{u}^{m(0)}_{s}  
\, {\cal P}_R \, \tilde{\chi}^0_l +h.c. \nonumber \\
&&  (i,j,k=1,2,3,\; s=1,..,6) \;. 
\eeq
Note, that $\tilde{W}^{(0)\dagger}_{is}\equiv\tilde{W}^{(0)\dagger}_{L
  is}$, $\tilde{W}^{(0) \dagger}_{j+3 \, s}\equiv \tilde{W}^{(0)
  \dagger}_{Rjs}$, {\it cf.} Eq.~(\ref{eq:sumw}).
Upon renormalization we replace \cite{Yamada:2001px}
\beq
\bar{u}^{m(0)} U^{u_L (0)} &\to& \bar{u}^{m} \left( 1 +
  \frac{\delta Z^{L\dagger}}{2} \right) (1 + \delta u^{u_L})  U^{u_L} 
\\
\tilde{W}^{(0) \dagger}_{L,R} \tilde{u}^{m(0)} &\to&
\tilde{W}^\dagger_{L,R} (1+\delta \tilde{w}^\dagger) \left( 1 +
\frac{\delta Z^{\tilde{u}}}{2} \right)\tilde{u}^m \\
{\bf m}^{(0)} &\to& {\bf m} + \delta {\bf m} \;,
\eeq
where we have suppressed the indices.
The wave function $\tilde{u}^m$ denotes a six-component column vector.
With the replacement $\tilde{W}_{L,R}
= W_{L,R} \,U^{u_{L,R}}$, {\it cf.} Eq.~(\ref{eq:factor}), we have
for the Yukawa part of the coupling
\beq
\bar{u}^{m} \left( 1 +
  \frac{\delta Z^{L\dagger}}{2} \right) (1 + \delta u^{u_L})  U^{u_L} 
({\bf m} + \delta {\bf m}) U^{u_R\dagger}  (1 + \delta u^{u_R\dagger})
W^\dagger_R (1+\delta \tilde{w}^\dagger)
\left( 1 + \frac{\delta Z^{\tilde{u}}}{2} \right) \tilde{u}^m \;,
\eeq
times $(-g Z_{l4})/(\sqrt{2}M_W s_\beta)$.
For the mass renormalization we choose the renormalization
prescription such that the bare mass matrices and hence $\delta {\bf
  m}$ are diagonal, {\it i.e.}
\beq
(1 + \delta u^{u_L})  U^{u_L} 
({\bf m} + \delta {\bf m}) U^{u_R\dagger}  (1 + \delta u^{u_R\dagger})
= ({\bf m}^D + \delta {\bf m}^D) \;,
\eeq
where $D$ denotes diagonal matrices. This is possible since the off-diagonal
elements can be absorbed into the off-diagonal elements of the
antihermitian part of the right-handed 
wave function renormalization matrices \cite{Balzereit:1998id}.
Exploiting the unitary of the mixing matrices we finally find for the
renormalized Lagrangian in the mass eigenstate basis
\beq
{\cal L}_{\bar{u}\tilde{u}\tilde{\chi}^0} &=& \bar{u}^m_i \,
(G^R_{isl}+\delta G^R_{isl}) \, {\cal P}_R \,\tilde{u}^m_{s} \tilde{\chi}^0_l +
\bar{u}^m_i (G^L_{isl}+\delta G^L_{isl}) \, {\cal P}_L \,
\tilde{u}^m_{s} \tilde{\chi}^0_l 
+ h.c. \;,
\eeq 
with the couplings given by
\beq
G^R_{isl} &=& -g e^{u_i}_{Ll} (W^\dagger_L)_{is} - \frac{g Z_{l4} m_{u_i}
  \delta_{ij}}{\sqrt{2} M_W\sin\beta} (W^\dagger_R)_{js} \\
G^L_{isl} &=& -g e^{u_i}_{Rl} (W^\dagger_R)_{is} - \frac{g Z_{l4} m_{u_i}
  \delta_{ij}}{\sqrt{2} M_W\sin\beta} (W^\dagger_L)_{js} \\
\delta G^R_{isl} &=& -g e^{u_i}_{Ll} \left[ \frac{\delta
    Z^{L\dagger}_{ij}}{2} (W^\dagger_L)_{js} + (W^\dagger_L)_{it} \frac{\delta
  Z^{\tilde{u}}_{ts}}{2} + \delta u^{u_L}_{ij} (W^\dagger _L)_{js} +
  (W^\dagger_L)_{it} \delta \tilde{w}^\dagger_{ts} \right] \nonumber\\
&& - \frac{g Z_{l4}}{\sqrt{2} M_W\sin\beta} \left[ \frac{\delta
    Z^{L\dagger}_{ij}}{2} m_{u_j} \delta_{jk} (W^\dagger_R)_{ks} + 
m_{u_i} \delta_{ij} (W^\dagger_R)_{jt} \frac{\delta
  Z^{\tilde{u}}_{ts}}{2} \right. \nonumber \\
&& \left. \hspace*{2.5cm} + m_{u_i} \delta_{ij} (W^\dagger_R)_{jt}
\delta \tilde{w}^\dagger_{ts} + \delta m_{u_i} \delta_{ij}
(W^\dagger_R)_{js}\right]  
\eeq
\beq
\delta G^L_{isl} &=& -g e^{u_i}_{Rl} \left[ \frac{\delta
    Z^{R\dagger}_{ij}}{2} (W^\dagger_R)_{js} + (W^\dagger_R)_{it} \frac{\delta
  Z^{\tilde{u}}_{ts}}{2} + \delta u^{u_R}_{ij} (W^\dagger _R)_{js} +
  (W^\dagger_R)_{it} \delta \tilde{w}^\dagger_{ts} \right] \nonumber\\
&& - \frac{g Z_{l4}}{\sqrt{2} M_W\sin\beta} \left[ \frac{\delta
    Z^{R\dagger}_{ij}}{2} m_{u_j} \delta_{jk} (W^\dagger_L)_{ks} + 
m_{u_i} \delta_{ij} (W^\dagger_L)_{jt} \frac{\delta
  Z^{\tilde{u}}_{ts}}{2} \right. \nonumber\\
&& \left. \hspace*{2.5cm} + m_{u_i} \delta_{ij} (W^\dagger_L)_{jt}
\delta \tilde{w}^\dagger_{ts} + \delta m_{u_i} \delta_{ij}
(W^\dagger_L)_{js}\right] \;. 
\eeq
In the framework of MFV at $\mu_{\mbox{\scriptsize MFV}}$ the $W$
matrix is diagonal in flavour space at 
tree level. At one-loop flavour off-diagonal elements are induced
through the wave function and the mixing matrix renormalization. 

\section*{Acknowledgments}

\noindent
We greatly acknowledge helpful discussions with Ben Allanach, Andreas
Crivellin, Manuel Drees, Bastian Feigl, Jaume Guasch, Jong
Soo Kim, Ulrich Nierste, Werner Porod, Heidi Rzehak, Pietro Slavich and
Michael Spira. We are grateful to Ulrich Nierste, Pietro Slavich and Michael Spira
for the careful reading of the manuscript. This research was supported
in part by the Deutsche Forschungsgemeinschaft via the
Sonderforschungsbereich/Transregio SFB/TR-9 Computational Particle
Phy\-sics. EP gratefully acknowledges support of the Graduiertenkolleg
``High Energy Physics and Particle  Astrophysics''.

\end{document}